\documentclass[sigconf,pbalance]{acmart}

\usepackage{graphicx}
\usepackage{soul}
\usepackage{flushend}
\usepackage[skip=0pt]{caption}
\usepackage{subcaption}
\usepackage{rotating}
\usepackage{xspace}
\usepackage{color, colortbl}
\usepackage{xcolor}
\usepackage{framed}
\usepackage{minted}
\usepackage{multirow}
\usepackage{enumitem}
\usepackage{listings}
\definecolor{codepurple}{RGB}{170,0,255}
\definecolor{codegreen}{RGB}{0,128,0}
\definecolor{codegray}{RGB}{110,110,110}
\definecolor{codeblue}{RGB}{0,92,197}
\definecolor{highlightlavender}{RGB}{245,237,250}

\lstdefinestyle{minted-java}{
    language=Java,
    basicstyle=\ttfamily\footnotesize,
    keywordstyle=\color{codegreen}\bfseries,
    identifierstyle=\color{black},
    stringstyle=\color{codegreen},
    commentstyle=\color{codegray},
    emph={getUsername,getPassword,editAccount,signon,setAccount,getAccount},
    emphstyle=\color{codeblue},
    moredelim=[is][\color{codepurple}]{@}{ },
    numbers=none,
    columns=fullflexible,
    showstringspaces=false,
    breaklines=true,
}

\newcommand{\tool}{\textsc{saint}\xspace}

\definecolor{problemblue}{RGB}{100,134,158}
\definecolor{idiomsgreen}{RGB}{0,162,0}
\definecolor{exercisebgblue}{rgb}{0,  .69,  .941}
\definecolor{deepgreen}{rgb}{0.0, 0.5, 0.0}
\definecolor{codegreen}{rgb}{0,0.6,0}
\definecolor{codegray}{rgb}{0.5,0.5,0.5}
\definecolor{codepurple}{rgb}{0.58,0,0.82}
\definecolor{backcolour}{rgb}{0.95,0.95,0.92}
\definecolor{redColor}{RGB}{255,0,0}
\definecolor{Gray}{gray}{0.1}
\definecolor{javared}{rgb}{0.6,0,0} % for strings
\definecolor{javagreen}{rgb}{0.25,0.5,0.35} % comments
\definecolor{javapurple}{rgb}{0.5,0,0.35} % keywords
\definecolor{javadocblue}{rgb}{0.25,0.35,0.75} % javadoc
\definecolor{ibmblue}{RGB}{63,97,246}
\definecolor{maroon}{RGB}{105,33,61}
\definecolor{teal}{RGB}{59,115,115}
\definecolor{exercisebgblue}{rgb}{0,  .69,  .941}

\definecolor{shadecolor}{RGB}{220,230,250}

\newenvironment{findingbox}[1]{%
  \vspace{-2pt}
  \begin{shaded}
  \noindent\textbf{Finding #1:}\ %
}{%
  \end{shaded}
  \vspace{-2pt}
}

\definecolor{gray05}{gray}{0.95}
\definecolor{gray08}{gray}{0.92}
\definecolor{gray10}{gray}{0.90}
\definecolor{gray12}{gray}{0.88}
\definecolor{gray15}{gray}{0.85}
\definecolor{gray18}{gray}{0.82}
\definecolor{gray20}{gray}{0.80}
\definecolor{gray25}{gray}{0.75}
\definecolor{gray30}{gray}{0.70}
\definecolor{gray35}{gray}{0.65}
\definecolor{gray40}{gray}{0.60}
\definecolor{gray45}{gray}{0.55}
\definecolor{gray50}{gray}{0.50}
\definecolor{gray55}{gray}{0.45}
\definecolor{gray60}{gray}{0.40}
\definecolor{gray65}{gray}{0.35}
\definecolor{gray70}{gray}{0.30}
\definecolor{gray75}{gray}{0.25}
\definecolor{gray80}{gray}{0.20}
\definecolor{gray85}{gray}{0.15}
\definecolor{gray90}{gray}{0.10}
\definecolor{gray95}{gray}{0.05}
\definecolor{blue}{HTML}{0f62fe}
\definecolor{red}{RGB}{234,51,35}
\definecolor{ibm-blue-light}{HTML}{a6c8ff}
\newcommand{\hlpink}[1]{{\setlength\fboxsep{1pt}\colorbox{magenta!12}{#1}}}
\newcommand{\hlblue}[1]{{\setlength\fboxsep{1pt}\colorbox{blue!12}{#1}}}

\newcommand{\filledcircle}[2]{% #1=color, #2=text
  \raisebox{-.4ex}{%
    \setlength{\fboxsep}{1pt}%
    \setlength{\fboxrule}{0.4pt}%
    \fcolorbox{black}{#1}{%
      \textcolor{white}{\small\bfseries #2}%
    }%
  }%
}
\newcommand{\bluelabeled}[2]{%
  \leavevmode
  \hbox{%
    \leaders\hbox{\rule{.4pt}{.4pt}\hskip1.2pt}\hskip #1ex
    \filledcircle{blue!90}{#2}%
  }%
}

\newcommand{\redlabeled}[2]{%
  \leavevmode
  \hbox{%
    \leaders\hbox{\rule{.4pt}{.4pt}\hskip1.2pt}\hskip #1ex
    \filledcircle{magenta!80}{#2}%
  }%
}

\newcommand{\smalltt}[1]{\texttt{\small #1}}

\newcommand{\urlpath}[1]{%
\mintinline[style=xcode]{yaml}
{#1}%
}

\setitemize{noitemsep,topsep=0pt,parsep=0pt,partopsep=0pt,leftmargin=*, wide=0pt}
\setenumerate{noitemsep,topsep=0pt,parsep=0pt,partopsep=0pt,leftmargin=*, wide=0pt}
\newcommand\bi{\begin{itemize}}
\newcommand\ei{\end{itemize}}
\newcommand\be{\begin{enumerate}}
\newcommand\ee{\end{enumerate}}

\newcommand{\bluecircle}[1]{%
  \raisebox{0.2ex}{%
    \setlength{\fboxsep}{1pt}%
    \setlength{\fboxrule}{0.4pt}%
    \fcolorbox{black}{blue!90}{%
      \textcolor{white}{\bfseries\small #1}%
    }%
  }%
}

\newcommand{\redcircle}[1]{%
  \raisebox{0.2ex}{%
    \setlength{\fboxsep}{1pt}%
    \setlength{\fboxrule}{0.4pt}%
    \fcolorbox{black}{magenta!80}{%
      \textcolor{white}{\bfseries\small #1}%
    }%
  }%
}

\usepackage{boldline}

\setminted{
    frame=lines,
    framesep=1mm,
    fontsize=\footnotesize,     % compact code font
    breaklines=true,
    breakanywhere=true,
    linenos=false,
    numbersep=0pt,
    escapeinside=||,
    mathescape=true,
    bgcolor=white,
    highlightcolor=magenta!12   % default highlight color
}

\usepackage{xparse} % in your preamble if not already

\NewDocumentEnvironment{tightminted}{O{}m} % [options]{language}
  {\vspace{-0.75\baselineskip}%
   \begin{minted}[#1]{#2}}
  {\end{minted}%
   \vspace{-0.5\baselineskip}}

\graphicspath{{./figures/}}
\AtBeginDocument{%
  }

\copyrightyear{2026}
\acmYear{2026}
\setcopyright{cc}
\setcctype{by-nc-nd}
\acmConference[ICSE '26]{2026 IEEE/ACM 48th International Conference on Software Engineering}{April 12--18, 2026}{Rio de Janeiro, Brazil}
\acmBooktitle{2026 IEEE/ACM 48th International Conference on Software Engineering (ICSE '26), April 12--18, 2026, Rio de Janeiro, Brazil}
\acmPrice{}
\acmDOI{10.1145/3744916.3773185}
\acmISBN{979-8-4007-2025-3/2026/04}
\begin{document}

% \title{SAINT: Generating Service-level Integration Tests using Static Analysis and LLMs}
\title{SAINT: Service-level Integration Test Generation with Program Analysis and LLM-based Agents}

\author{Rangeet Pan}
\email{rangeet.pan@ibm.com}

\affiliation{%
  \institution{IBM Research}
  \city{Yorktown Heights}
  \state{NY}
  \country{USA}
}

\author{Raju Pavuluri}
\email{pavuluri@us.ibm.com}

\affiliation{%
  \institution{IBM Research}
  \city{Yorktown Heights}
  \state{NY}
  \country{USA}
}

\author{Ruikai Huang}
\email{rkh@gatech.edu}
\affiliation{%
  \institution{Georgia Institute of Technology}
  \city{Atlanta}
  \state{GA}
  \country{USA}
}

\author{Rahul Krishna}
\email{rkrsn@ibm.com}

\affiliation{%
  \institution{IBM Research}
  \city{Yorktown Heights}
  \state{NY}
  \country{USA}
}
\author{Tyler Stennett}
\authornote{Author was an intern at IBM Research at the time of this work.}
\email{tstennett3@gatech.edu}
\affiliation{%
  \institution{Georgia Institute of Technology}
  \city{Atlanta}
  \state{GA}
  \country{USA}
}

\author{Alessandro Orso}
\email{orso@uga.edu}
\affiliation{%
  \institution{University of Georgia}
  \city{Athens}
  \state{GA}
  \country{USA}
}

\author{Saurabh Sinha}
\email{sinhas@us.ibm.com}
\affiliation{%
  \institution{IBM Research}
  \city{Yorktown Heights}
  \state{NY}
  \country{USA}
}

\renewcommand{\shortauthors}{Pan et al.}

\begin{abstract}
Enterprise applications are typically tested at multiple levels, with service-level testing playing an important role in validating application functionality. Existing service-level testing tools, especially for RESTful APIs, often employ fuzzing and/or depend on OpenAPI specifications which are not readily available in real-world enterprise codebases. Moreover, these tools are limited in their ability to generate functional tests that effectively exercise meaningful scenarios. In this work, we present \tool, a novel white-box testing approach for service-level testing of enterprise Java applications. \tool combines static analysis, large language models (LLMs), and LLM-based agents to automatically generate endpoint and scenario-based tests. The approach builds two key models: an endpoint model, capturing syntactic and semantic information about service endpoints, and an operation dependency graph, capturing inter-endpoint ordering constraints.
\tool then employs LLM-based agents to generate tests. Endpoint-focused tests aim to maximize code and database interaction coverage. Scenario-based tests are synthesized by extracting application use cases from code and refining them into executable tests via planning, action, and reflection phases of the agentic loop. We evaluated \tool on eight Java applications, including a proprietary enterprise application. Our results illustrate the effectiveness of \tool in coverage, fault detection, and scenario generation. Moreover, a developer survey provides strong endorsement of the scenario-based tests generated by \tool. 
Overall, our work shows that combining static analysis with agentic LLM workflows enables more effective, functional, and developer-aligned service-level test generation.
\end{abstract}




\maketitle

\section{Introduction}
\label{sec:into}

\begin{figure}[t]
    \centering
    \includegraphics[width=0.86\linewidth]{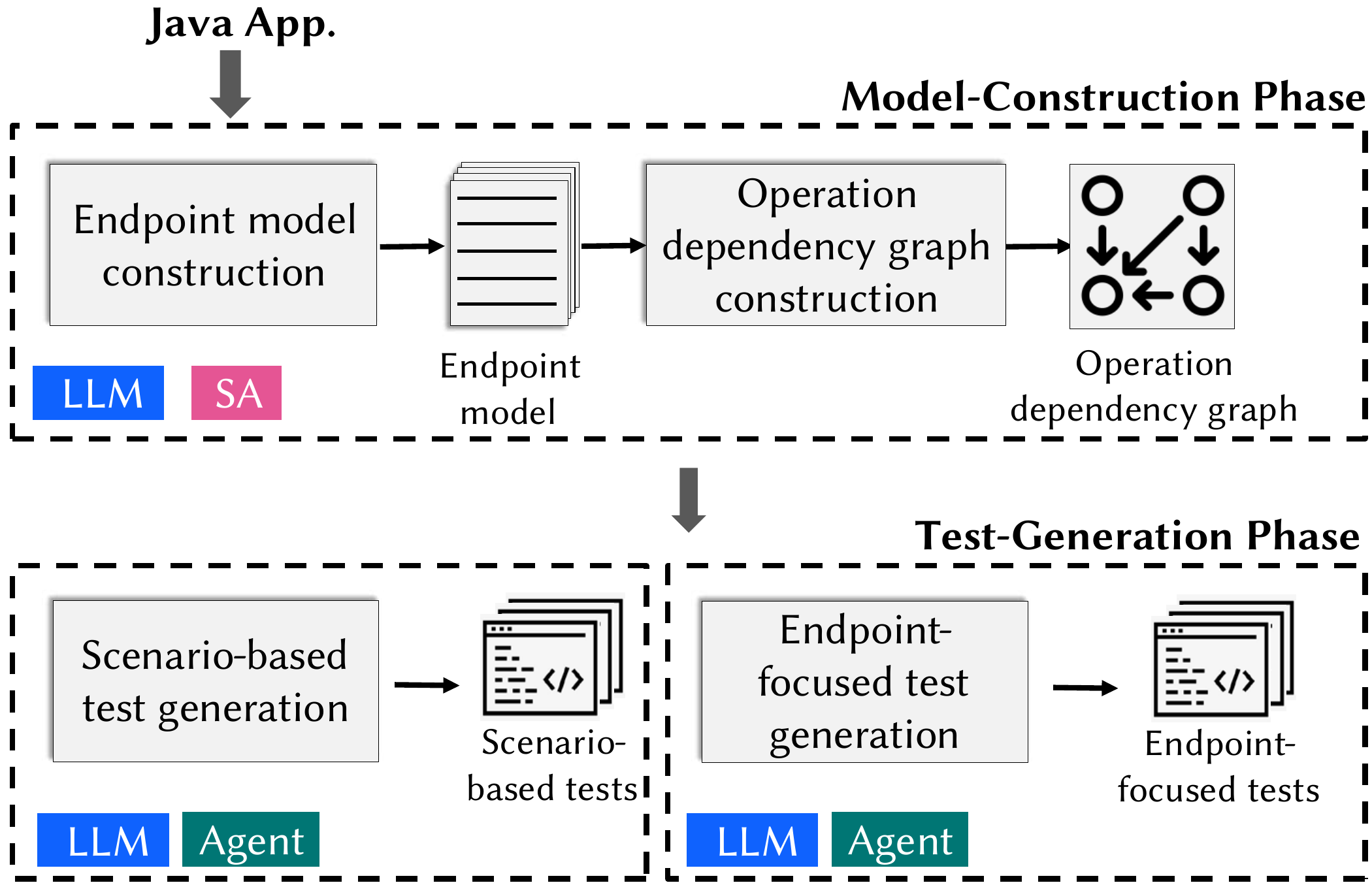}
    \vspace{4pt}
    \caption{Overview of our approach.} 
    % \vspace{-3pt}
    \label{fig:overview}
\end{figure}

Enterprise applications are large, multi-tiered systems with complex business logic. To gain confidence in their correctness, such applications are tested at multiple levels, each focusing on different validation goals. Unit testing checks individual implementation units (e.g., methods, functions, or classes) in isolation, whereas end-to-end testing exercises functional flows across application tiers. Between these levels, service-level testing works at the service layer of the application, with the goal of validating the service endpoints and server-side logic. It is typically guided by coverage goals over service operations, their parameters, and reachable code, as well as operation sequences that reflect functional flows or use cases. 
% <--Change for CRC
\begin{figure*}[t!]
\begin{minipage}[t]{0.5\linewidth}
% \vspace{-0.5em}
\begin{minted}[baselinestretch=0.65, fontsize=\scriptsize, highlightlines={6, 10,14,19},highlightcolor=blue!12]{java}
@PostMapping("/owners/{ownerId}/pets/{petId}/visits/new")
public String processNewVisitForm(@ModelAttribute Owner owner, 
    @ModelAttribute Pet pet, @Valid Visit visit, BindingResult result) {
    
    // Cross-path parameter relationship
    if |{(pet.getId() != owner.getId())}\bluelabeled{34.4}{1}| 
        throw new IllegalArgumentException("Pet and owner mismatch");
    
    // Temporal constraint validation
    if |\hlblue{(visit.getDate().isBefore(LocalDate.now()))}\bluelabeled{19.5}{2}| 
        result.rejectValue("date", "past.date");
   
   // Context specific validation
   if |\hlblue{(isEmergency(visit.getDescription())} \bluelabeled{27}{3}|
        if(visit.getDate().isAfter(LocalDate.now().addDays(1)))
            result.rejectValue("date", "not.same.day");
    
    // Multiparameter conditional validation
    if |\hlblue{(isSurgeryVisit(visit.getDescription())) \{}\bluelabeled{20.5}{4}|
        if (visit.getDate().isBefore(LocalDate.now().addDays(7)))
            result.rejectValue("date", "surgery.advance.booking");
        if (visit.getDescription.length() < 50) 
            result.rejectValue("description", "detail.required");
    }
    if (result.hasErrors()) { return "pets/createOrUpdateVisitForm"; }
    return "redirect:/owners/{ownerId}"; }
\end{minted}
% \vspace{-15pt}
\captionof{listing}{{Real-world endpoint from PetClinic with hard-to\\-test constraints.}}
\label{lst:motivation-example-1}
\end{minipage}%
\begin{minipage}[t]{0.5\linewidth}
% \vspace{-0.5em}
\begin{minted}[escapeinside=||, frame=lines,framesep=1mm,baselinestretch=0.4, fontsize=\scriptsize, breaklines, breakanywhere, highlightlines={4,12,17,23,24,25,26},highlightcolor=magenta!12]{java}
@Test public void testCreateAndUpdatePetForAnOwner() {
    // Initialize creation form for a new pet
    Response response = given()
            .when().get("/owners/1/pets/new")|\redlabeled{20.2}{1}|
            .then().statusCode(200).extract().response();

    // Process creation form for a new pet for owner
    response = given()
            .contentType("application/x-www-form-urlencoded")
            .formParam("pet.name", "Buddy")
            .formParam("pet.birthDate", "2023-01-01")
            .when().post("/owners/1/pets/new")|\redlabeled{19.4}{1}|
            .then().statusCode(200).extract().response();

    // Initialize update form for the pet of owner
    response = given().when()
            .get("/owners/1/pets/1/edit")|\redlabeled{25}{2}|
            .then().statusCode(200).extract().response();

    // Process update form to change date of birth
    response = given()
            .contentType("application/x-www-form-urlencoded")
            .formParam("pet.name", "Buddy")|\redlabeled{23}{3}|
            .formParam("pet.birthDate", "2022-01-01")|\redlabeled{12.1}{3}|
            .when().post("/owners/1/pets/1/edit")|\redlabeled{16.5}{2}|
            .then().statusCode(200).extract().response();|\redlabeled{7.8}{4}|}
\end{minted}
% \vspace{-15pt}
\captionof{listing}{A test synthesized by \tool capturing these constraints in a realistic scenario.}
\label{lst:motivating-example-2}
\end{minipage}
\vspace{-11pt}
\end{figure*}

In this work, we focus on improving service-level (or API-level) testing of enterprise Java applications. Our goal is two-fold: to generate high-coverage tests for individual service operations or endpoints, and to create scenario-based tests that exercise sequences of operations corresponding to coherent use cases. Although many service-level testing techniques exist—targeting RESTful APIs~\cite{atlidakis_restler_2019, arcuri2019restful, viglianisi_resttestgen_2020, kim2023reinforcement, kim2023enhancing, kim_llamaresttest_2025, kim_multi-agent_2025, corradini2024deeprest, saha:2025:raft, liu_morest_2022, martin2021restest, martin2022online, martin2021:blackandwhite, zhang_logiagent_2025}, GraphQL APIs (e.g.,~\cite{belhadi:2024:graphql, karlsson:2020:graphql}) and the older WSDL-based services (e.g.~\cite{bartolini:2009:wstaxi})—they have key limitations that restrict their applicability for functional test generation on enterprise Java applications. First, most REST API testing tools function as fuzzers and do not produce test cases. Second, techniques that generate tests focus on maximizing code coverage~\cite{arcuri2019restful} or infer sequences from producer-consumer or resource-based dependencies~\cite{saha:2025:raft}; however, meaningful operation sequences can exist without such dependencies. Finally, most approaches rely on formal service specifications—typically OpenAPI~\cite{openapi:spec}—which are often not available for enterprise applications.

We present a new white-box technique, called \tool, for service-level testing that combines static program analysis with the power of large language models (LLMs)---leveraging their planning, reasoning, and reflection capabilities through agentic workflows to enhance test generation effectiveness. Our approach (shown in Figure~\ref{fig:overview}) generates \textit{endpoint-focused tests} that exercise each service endpoint with the goal of maximizing code coverage, and \textit{scenario-based tests} that focus on covering meaningful use cases of the application for functional testing. The approach consists of two phases: model-construction and test-generation. 

The model-construction phase analyzes the application under test to identify service endpoints of the application and constructs an \textit{endpoint model} that incorporates (1)~syntactic information for an endpoint that consists of endpoint path, parameter names, and parameter types, and (2)~semantic information for an endpoint with inter-parameter dependencies~\cite{martin:2019:catalog} and parameter value constraints. Additionally, we construct an \textit{operation dependency graph} (ODG) that captures ordering constraints between endpoints resulting from different types of dependencies. Both these models are constructed using a combination of static analysis and carefully crafted LLM prompts, with suitable in-context learning examples.

The test-generation phase of \tool uses the endpoint model and the ODG to create two test types based on user intent: endpoint-focused tests and scenario-based tests.
The endpoint-focused tests explore each service endpoint with varied inputs to maximize code coverage, with emphasis on covering database interaction points in code. To create these tests, \tool constructs an LLM prompt for each endpoint using syntactic, semantic, and ordering constraints from the ODG, sends it to the LLM, and extracts parameter value sets from the received LLM response. It then builds concrete HTTP requests and executes them against the deployed APIs, while monitoring coverage of  application code. \tool incorporates a \textit{repair agent} that attempts to fix invalid requests (i.e., requests for which the server returns 4xx response codes) and a \textit{coverage-augmentation agent} that attempts to increase coverage of reachable code for an endpoint; both these agents implement an iterative planning, action execution (with an available set of tools), and reflection workflow.

% and builds HTTP requests to test APIs while monitoring coverage. \tool enhances tests with two agents: a \textit{repair agent} to fix invalid requests (4xx errors) and a \textit{coverage-augmentation agent} to increase endpoint code coverage; both use iterative planning, execution, and reflection. 

To create scenario-based tests, which exercise meaningful application use cases, \tool performs a sequence of LLM prompting to first extract test scenarios (in Gherkin-like syntax~\cite{gherkin:syntax}) from the application code and then refine the scenarios into atomic blocks or test steps. These atomic blocks are input to a \textit{test-generation agent} that attempts to reify a test scenario into an executable test case. This agent, like the repair and coverage-augmentation agents, implements a plan-act-reflect loop, with a suitable set of tools for executing actions. The output of the agent consists of Java test code fragments corresponding to the atomic blocks, which are then assembled, via another LLM call, into the final scenario-based test.

% (2)~\textit{scenario-based tests:} tests application use cases, with \tool extracting test scenarios (in Gherkin-like syntax~\cite{gherkin:syntax}) from the code, then refining them into atomic blocks. These blocks serve as input for a \textit{test-generation agent} that converts them into executable test cases, using a plan-act-reflect loop and specific tools. The result is Java test code fragments assembled via another LLM call into the final scenario-based test.

We evaluated \tool on eight Java applications, including a proprietary enterprise application. Four of these applications have OpenAPI specifications; for these applications, we compared \tool against EvoMaster, a state-of-the-art white-box test generation tool for RESTful APIs~\cite{arcuri2019restful}. We assessed \tool's effectiveness through coverage metrics and fault detection by server failure measurement. We evaluated the scenario-based tests for coverage, characteristics of the extracted scenarios, and developer perception, collected via a user survey. Finally, we analyzed \tool's key components through an ablation study.
Our results show that \tool matches or considerably outperforms EvoMaster on code coverage achieved with the endpoint-focused tests, but is not as effective as EvoMaster in triggering server failures, with scope for improvement.
In terms of scenario-based tests, \tool effectively extracts scenarios that span multiple endpoints and are rated highly by developers, with more than 90\% of the survey participants stating that they would test application scenarios similar to the extracted ones. The ablation study highlights the contributions of \tool's key components. %, including ODG construction and semantic information extraction.

The main contributions of the work are:

\begin{itemize}[leftmargin=*]

\item A novel technique that combines static analysis with LLM prompting and agentic workflows to generate endpoint-focused and scenario-based service-level tests that aim to maximize coverage while also exercising meaningful use cases for functional testing.
\item Empirical results showing \tool's effectiveness in code coverage, fault detection, and scenario extraction, with developer feedback highlighting the value of \tool-generated scenario-based tests.
\item An artifact consisting of experiment scripts, data, and LLM prompts that is publicly available~\cite{supplementary}.
\end{itemize}

% \rangeet{We have made the results and the prompts available at~\cite{supplementary}.}

\section{Motivation}
\label{sec:motivation}

In this section, we ground our discussion by presenting a reference example from Spring PetClinic~\cite{spring-petclinic}, a multi-tier application that exposes multiple service endpoints requiring coordinated inputs, business rule enforcement across entities, and dynamic state-dependent database access. This example highlights some of the challenges for test generation. %how traditional testing will fail, especially when faced with undocumented logic, shared state, and indirect effects across service layer.
We then present an example test case generated by our tool, demonstrating how it captures a realistic multi-endpoint sequence and validates non-trivial state-dependent behaviors.

% <--CRC change
\noindent\textbf{Service-level testing challenges for a PetClinic endpoint.}~%
We consider the \urlpath{.../{ownerId}/.../{petId}/visits/new} endpoint from PetClinic, shown in Listing~\ref{lst:motivation-example-1}. This example illustrates several challenges: \bluecircle{1} a value constraint on the \smalltt{owner} parameter to ensure the provided \smalltt{petId} belongs to the specified \smalltt{ownerId}; \bluecircle{2} a temporal check to assert that the visit date is not in the past; \bluecircle{3} an implicit dependency that exists between the description text and the date (i.e., emergency visits must be within a day); \bluecircle{4} a condition that ties the surgery type to the date and the description.
All the above clauses span entity relations, involving various inter-parameter dependencies, which cannot be adequately captured or represented using OpenAPI specifications or static analysis alone. 
% We ought to look beyond conventional spec-driven test generation approaches to adequately test these endpoints.

% %%%%%%%%%%%%%%%%%%%%%%%%%%%%%%%%%%%%%[ OUR EXAMPLE ]%%%%%%%%%%%%%%%%%%%%%%%%%%%%%%%%%%%%%%%%%%%%%%%%%%%%%%
\noindent\textbf{An illustrative test synthesized by our approach.}~
The test shown in Listing~\ref{lst:motivating-example-2} demonstrates how our approach addresses the challenges in scenario-based testing. It constructs a coherent multi-endpoint test that maintains semantic consistency across endpoints with consistent use of owner ID 1 across the test (as seen in \redcircle{1}). It also enforces state-dependent constraints (as seen in \redcircle{2}, where the edit operations assume that the pet was successfully created beforehand), and embeds realistic validation logic (e.g., via valid form data entries on lines labeled~\redcircle{3}).  In summary, the example exposes cross-endpoint dependencies and produces a test case with valid inputs and assertions (e.g., final check on line~\redcircle{4}), illustrating how our approach can perform effective service-level test generation.

\section{Methodology}
\label{sec:methodology}

% \begin{figure}[t]
%     \centering
%     \includegraphics[width=\linewidth]{figures/saint-overview.pdf}
%     % \vspace{-20pt}
%     \caption{Overview of \tool. \hl{This figure is too small. Also, it may be better to put this in page 1 second column in the intro?}} 
%     % \vspace{-5pt}
%     \label{fig:overview}
% \end{figure}

% \input{model/endpoint_path}
% \input{model/database}
% \input{model/endpoint_parameter}
% \input{model/ipd}

\tool operates % over a Java application (with a service layer)
in two phases where Phase~1 constructs the endpoint model and the ODG and Phase~2 performs endpoint-based and scenario-based test generation.

\subsection{Endpoint Model Construction}

\begin{figure}[t]
    \centering
    \includegraphics[width=.86\linewidth]{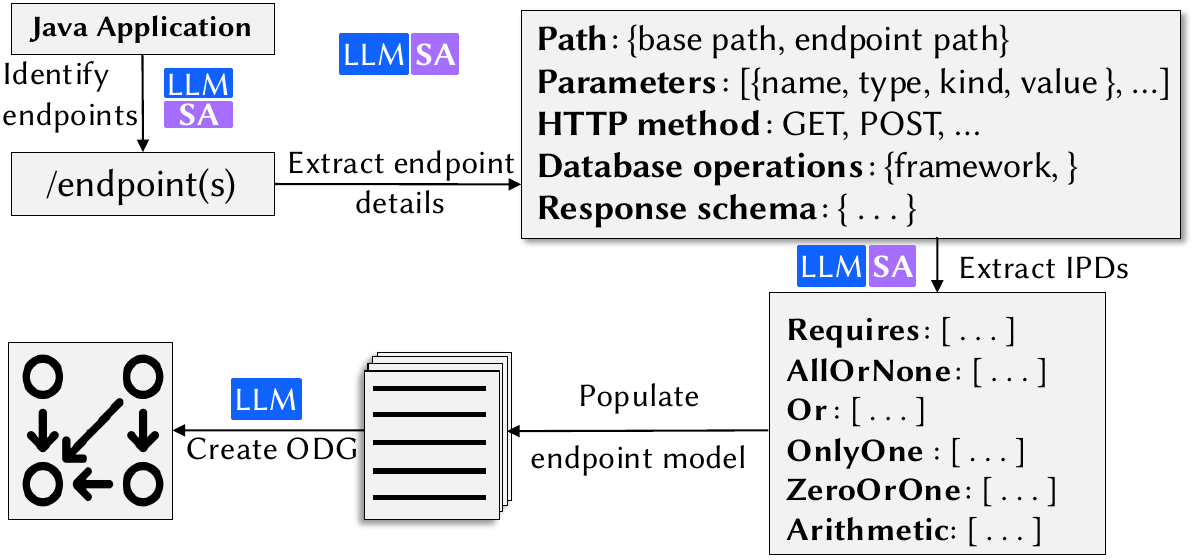}
    \vspace{2pt}
    \caption{Construction of the endpoint model and ODG via static analysis (SA) and LLM prompting (LLM).} 
    % \vspace{-3pt}
    \label{fig:endpoint_model}
\end{figure}

Figure~\ref{fig:endpoint_model} illustrates the construction our endpoint model using static analysis and LLM prompting. First, we identify endpoint methods in the application written in various Java frameworks. \tool currently supports five Java frameworks: Jakarta~\cite{jakarta}, Spring~\cite{spring}, Struts~\cite{struts}, Stripes~\cite{stripes}, and JDK HttpServer~\cite{jdkhttpserver}. For Jakarta, Spring, and Struts, we rely on CLDK ~\cite{cldk}. CLDK is a multi-language program analysis library that is built upon well-vetted static analysis tools, such as WALA~\cite{wala}, JavaParser~\cite{javaparser}, and Tree-sitter~\cite{treesitter}, and provides various analysis capabilities via Python APIs. We use these APIs to extract symbol table, call graph, and endpoint and database information and build custom analysis for this work. To identify endpoints, CLDK relies on a pattern-matching technique that was proposed in a previous work~\cite{antoniadis2020static}. For Stripes and HttpServer, \tool implements custom static analyses and LLM prompts to identify endpoints. These analyses leverage the symbol table information (e.g., method implementations, annotations, parameter details, etc.) and the call graph (constructed with Rapid Type Analysis~\cite{bacon1996fast}) obtained using the CLDK APIs. After identifying the endpoints, \tool extracts detailed syntactic and semantic information for each endpoint to populate the endpoint model. %\rangeet{For most usecases, \tool leverage Rapid Type Analysis-based~\cite{bacon1996fast} call-graph analysis and symbol table information such as code body, annotations, parameter details, calling type, etc.}

\begin{figure}[t]
\centering
% Display the tuple expression
% \begin{minipage}[t]{\linewidth}
% Resize table to fit column width
\resizebox{0.86\linewidth}{!}{%
% %\renewcommand{\arraystretch}{1.25}
\begin{tabular}{ccl}
\hlineB{2}
\multicolumn{1}{V{2}c}{\textbf{Symbol}} & \multicolumn{1}{l}{\textbf{Type}}            & \multicolumn{1}{cV{2}}{\textbf{Description}}                                \\\hlineB{2}
\multicolumn{3}{c}{}                                                     \\[-1em]\hlineB{2}
\multicolumn{1}{V{2}c}{\cellcolor{gray08}} & \multicolumn{1}{l}{\cellcolor{gray08}} & \multicolumn{1}{lV{2}}{\cellcolor{gray08}{$\text{\textbf{API endpoint~}} \mathbf{(\mathcal{E}) \equiv (c, m, p, H, \Pi, I, D, R)}$}}\\
\multicolumn{3}{c}{}                                                     \\[-1.2em]
\multicolumn{1}{V{2}c}{$c$}    & \multicolumn{1}{l}{$\Sigma^+$}      & \multicolumn{1}{lV{2}}{Fully qualified name of the class containing the endpoint method.}                         \\
\multicolumn{1}{V{2}c}{$m$}    & \multicolumn{1}{l}{$\Sigma^+$}      & \multicolumn{1}{lV{2}}{Signature of the API endpoint method}                    \\

% \multicolumn{1}{V{2}c}{$p$}    & \multicolumn{1}{c{$\Sigma^*$}      & \multicolumn{1}{lV{2}}{Endpoint path}                                \\
\multicolumn{1}{V{2}c}{$p$}      & \multicolumn{1}{l}{$\Sigma^*$}  & \multicolumn{1}{lV{2}}{Endpoint path}         \\
\multicolumn{1}{V{2}c}{$H$}      & \multicolumn{1}{l}{$\mathbb{H}$}  & \multicolumn{1}{lV{2}}{HTTP method. $\mathbb{H}=\{GET, POST, PUT, DELETE, PATCH\}$}         \\

% \multicolumn{1}{V{2}c}{$\Pi$}  & \multicolumn{1}{l}{$\mathcal{P}^*$} & \multicolumn{1}{V{2}}{List of endpoint parameters}                  \\

\multicolumn{1}{V{2}c}{$\Pi$}    & \multicolumn{1}{l}{$\mathcal{P}^*$} & \multicolumn{1}{lV{2}}{List of endpoint parameters}         \\
\multicolumn{1}{V{2}c}{$I$}    & \multicolumn{1}{l}{$\mathcal{I}^*$} & \multicolumn{1}{lV{2}}{List of inter-parameter dependencies}         \\
\multicolumn{1}{V{2}c}{$D$}    & \multicolumn{1}{l}{$\mathcal{D}^*$} & \multicolumn{1}{lV{2}}{List of database operations}                  \\
\multicolumn{1}{V{2}c}{$R$}      & \multicolumn{1}{l}{$\Sigma^*$}    & \multicolumn{1}{lV{2}}{Response schema as dictionary or string reference}             \\ \hlineB{2}
\multicolumn{3}{c}{}\\[-0.95em]\hlineB{2}
\multicolumn{1}{V{2}c}{\cellcolor{gray08}} & \multicolumn{1}{l}{\cellcolor{gray08}} & \multicolumn{1}{lV{2}}{\cellcolor{gray08}{$\text{\textbf{Endpoint Parameter}~}\mathbf{(\mathcal{P}) \equiv (n, T, K, V, C, M, A)}$}}\\
\multicolumn{1}{V{2}c}{$n$}    & \multicolumn{1}{l}{$\Sigma^*$}      & \multicolumn{1}{lV{2}}{Name of the parameter}                        \\
\multicolumn{1}{V{2}c}{$T$}    & \multicolumn{1}{l}{$\Sigma^*$}      & \multicolumn{1}{lV{2}}{Datatype of the parameter}                    \\
\multicolumn{1}{V{2}c}{$K$}      & \multicolumn{1}{l}{$\mathcal{K}$} & \multicolumn{1}{lV{2}}{Parameter kind. $\mathcal{K}=\{query, path, body, header\}$}                 \\
\multicolumn{1}{V{2}c}{$V$}      & \multicolumn{1}{l}{$\Sigma^*$}    & \multicolumn{1}{lV{2}}{Value constraints (e.g., allowed strings, enum options)}             \\
\multicolumn{1}{V{2}c}{$M$}    & \multicolumn{1}{l}{$\Sigma^*$}      & \multicolumn{1}{lV{2}}{Enclosing method for the parameter}           \\
\multicolumn{1}{V{2}c}{$C$}    & \multicolumn{1}{l}{$\Sigma^*$}      & \multicolumn{1}{lV{2}}{Enclosing class for the parameter}            \\
\multicolumn{1}{V{2}c}{$A$}    & \multicolumn{1}{l}{$\Sigma^*$}      & \multicolumn{1}{lV{2}}{List of annotations applied to the parameter} \\ \hlineB{2}
\multicolumn{3}{c}{}\\[-0.95em]\hlineB{2}
\multicolumn{1}{V{2}c}{\cellcolor{gray08}} & \multicolumn{1}{l}{\cellcolor{gray08}} & \multicolumn{1}{lV{2}}{\cellcolor{gray08}{$\text{\textbf{Inter-parameter dependency }~}\mathbf{(\mathcal{I}) \equiv (R, \Pi, \Gamma)}$}}\\
\multicolumn{1}{V{2}c}{$R$}    & \multicolumn{1}{l}{$\mathbb{I}$}    & \multicolumn{1}{lV{2}}{Inter-parameter dependency relation type.}    \\
\multicolumn{1}{V{2}c}{$\Pi$}  & \multicolumn{1}{l}{$\mathcal{P}^+$} & \multicolumn{1}{lV{2}}{List of involved parameters}                  \\
\multicolumn{1}{V{2}c}{$\Gamma$} & \multicolumn{1}{l}{$\Sigma^*$}    & \multicolumn{1}{lV{2}}{Constraint logic associated with the relation}                       \\ \hlineB{2}
\multicolumn{3}{c}{}\\[-0.95em]\hlineB{2}
\multicolumn{1}{V{2}c}{\cellcolor{gray08}} & \multicolumn{1}{l}{\cellcolor{gray08}} & \multicolumn{1}{lV{2}}{\cellcolor{gray08}{$\text{\textbf{Database operation }~}\mathbf{(\mathcal{D}) \equiv (F, C, L, M, O)}$}}\\
\multicolumn{1}{V{2}c}{$F$}      & \multicolumn{1}{l}{$\mathcal{F}$} & \multicolumn{1}{lV{2}}{Database framework. $\mathcal{F}=\{JDBC, JPA, JTA, ...\}$}           \\
\multicolumn{1}{V{2}c}{$C$}    & \multicolumn{1}{l}{$\Sigma^*$}      & \multicolumn{1}{lV{2}}{Enclosing class name of the operation}        \\
\multicolumn{1}{V{2}c}{$M$}    & \multicolumn{1}{l}{$\Sigma^*$}      & \multicolumn{1}{lV{2}}{Method signature where the DB access occurs}  \\
\multicolumn{1}{V{2}c}{$L$}    & \multicolumn{1}{l}{$i$}             & \multicolumn{1}{lV{2}}{Line number where operation has been taken}   \\
\multicolumn{1}{V{2}c}{$O$}      & \multicolumn{1}{l}{$\mathcal{O}$} & \multicolumn{1}{lV{2}}{CRUD operation type. $\mathcal{O}=\{Create, Read, Update, Delete\}$} \\ \hlineB{2}
\end{tabular}
}
% \end{minipage}
\vspace{2pt}
\caption{The endpoint model constructed by \tool.}
\label{fig:endpoint-model}
\end{figure}

Figure~\ref{fig:endpoint-model} formally defines the endpoint model. An endpoint $\mathcal{E}$, corresponding to a method, is represented as an 8-tuple consisting of service class name $c$, method signature $m$, endpoint path $p$, HTTP method $H$, endpoint parameters $\Pi$, inter-parameter dependencies (IPDs)~\cite{martin:2019:catalog} $\mathcal{I}$, reachable database operations $\mathcal{D}$, and the response schema $\mathcal{R}$, representing the structure of the server response. The figure also shows the data type of each field in the model: a field is either a custom type defined within the model (e.g., $\mathbb{H}$ for HTTP methods), a string type ($\Sigma$), or an integer type ($i$).

% \subsubsection{Identifying endpoints}

\noindent\textbf{Extracting endpoint path and HTTP method.}~% For an endpoint, the resource path can declared at the class level, which is shared across all the endpoints in the class, or at the endpoint method level.
% 
% The path and operation type declaration styles vary in different Java frameworks. In most cases, they are declared as part of class or method-level annotations, which \tool extracts via static analysis. However, in legacy frameworks like HttpServer, path details are embedded in the service code. For instance, in the following code fragment from the LanguageTool service~\cite{languagetool}, paths must be extracted from conditional logic implemented in code. In such cases, we prompt an LLM to extract path information.
% 
Java frameworks use various patterns for declaring paths and operation types. Typically, these are class or method annotations that can be extracted via code parsing. However, in the case of HttpServer, a legacy framework that lacks annotation-based conventions, endpoint paths are specified in code, as shown in this example from LanguageTool~\cite{languagetool}: \\[-1.5em] 
%<--CRC Change
\begin{minted}[
    frame=lines, 
    baselinestretch=1, 
    highlightlines={2,4}, 
    highlightcolor=magenta!12, 
    style=vs
]{java}
void handleRequest(String path, ...) throws Exception {
    if (path.equals("languages")) {
      handleLanguagesRequest(httpExchange);
    } else if (path.equals("maxtextlength")) {...}
\end{minted}
% \begin{figure}[H]
%     \centering
%     \includegraphics[width=\linewidth]{figures/handlerequest.png}
%     \caption*{} % prevents increment
%     % \addtocounter{figure}{-1} % restores counter
% \end{figure}
% \vspace{-1.7em}
In this case, \tool\ relies on LLMs to extract the paths. Although complex static analysis could, in principle, support applications built on HttpServer, we deemed such an effort unwarranted due to the framework’s limited usage.

\noindent\textbf{Extracting endpoint parameter details.}~The parameter information in the model (Figure~\ref{fig:endpoint-model}) consists of the parameter name $n$, the parameter type $T$, the parameter kind $K$, value constraints $V$, enclosing method $M$ and class $C$, and the associated annotations $A$. 
% For each endpoint, we store all required details for each parameters. The extraction logic significantly varies based on the application framework and we discuss them in details in the following paragraphs.
% \vskip 2pt
% \noindent
% \textit{\underline{Parameter name and type:}}
\vskip 2pt
\noindent\textit{Parameter names and types.} In Spring and JAX-RS, parameters are declared in the endpoint method, allowing easy extraction of names and types (unless we encounter complex patterns \smalltt{@ModelAttribute} in Spring). In other cases, such as Jakarta Servlets, more advance processing is required as shown in the below snippet. Here, the parameter \smalltt{OrderProcessingMode} stored as a \smalltt{String} and then converted to \smalltt{int}. Here, \smalltt{getParameter()} calls on \smalltt{HttpRequest} are analyzed (see \hlpink{highlighted} lines) where the parameter values are \smalltt{String} typed (return type of \smalltt{getParameter()}) and are subsequently typecast to \smalltt{int}. These processes can happen anywhere in the call chain starting at the endpoint method.

%<--CRC change
\begin{minted}[highlightlines={2,4}]{java}
void doConfigUpdate(HttpServletRequest req) throws Exception {    
    String modeStr = req.getParameter("OrderProcessingMode");
    if (orderProcessingModeStr.isNotNull())
        int i = Integer.parseInt(modeStr);}// additional logic below
\end{minted}

% \begin{figure}[H]
%     \centering
%     \includegraphics[width=\linewidth]{figures/doconfigupdate.png}
%     \caption*{} % prevents increment
%     % \addtocounter{figure}{-1} % restores counter
% \end{figure}

\noindent To handle such cases, \tool performs a call-chain analysis, scoping it to the methods to which the \smalltt{HttpRequest} object flows via parameter passing. Within this scope, it identifies parameter names and types via LLM prompting, including relevant code fragments from the call graph and a simple in-context example in the prompt.
% 
% \tool handles these cases via LLM calls. The LLM prompt includes code fragments of methods in the call chain and the parameter names, and instructs the LLM to extract the type in a structured format, with a simple in-context example provided in the prompt. 
% 
% However, these information can be present in any method in the call chain starting from the endpoint method, where the \texttt{HttpRequest} object has been passed as an argument. Here, we extract the call graph and identify all the parameters for each endpoint. In contrast, frameworks like Spring simplify parameter identification, as parameters are directly associated with the endpoint method annotations. However, certain patterns, such as the use of the \texttt{@ModelAttribute} annotation, require additional processing. This annotation instructs the compiler to include all parameters from the method to be  accessible to all the endpoints in the same class, which warrants to process additional method(s) in the same endpoint class.
% 
% \paragraph{Parameter type ($T$)} To extract the type of a parameter, \tool relies on static analysis for all supported frameworks except Servlets. In Spring, parameter types can be inferred directly from formal parameters of the endpoint method. However, other frameworks require more intricate analysis.
% 

The following example from JPetStore~\cite{jpetstore} (a Stripes application~\cite{stripes}) illustrates how parameters can be declared as class fields linked via getter-setter methods. The lines \hlpink{highlighted} show how the \smalltt{editAccount()} method passes an \smalltt{account} object to \smalltt{setAccount()}. As \smalltt{AccountService} is annotated with \smalltt{@Service}, its fields become parameters. In the \smalltt{signon()} method, parameters are inferred from getters for \smalltt{username} and \smalltt{password}.\\[-1.6em]
%<--CRC change
\begin{minted}[highlightlines={6,7}]{java}
@Service public class AccountService {private Account account;}
public class AccountActionBean {
    private Account account = new Account();
    public String getUsername() { return account.getUsername(); }
    public String getPassword() { return account.getPassword(); }
    public Resolution editAccount() { accountService.setAccount(account); }
    public Resolution signon() {account = accountService.getAccount(getUsername(), getPassword()); }}
\end{minted}
\noindent\textit{Parameter kind.} Parameter kind can be path, query, header, or body, indicating whether the parameter is included in the resource path, the request query string, or the request body. \tool extracts this information using static analysis.

\vskip 2pt
\noindent\textit{Parameter value constraints.} Parameters can have value constraints enforced by code checks or annotations. In some frameworks, developers provide natural language examples to aid in generating corresponding values.\\
\vspace{-1.4em}
%<-- CRC change
\begin{minted}[highlightlines={3}]{java}
List<VariantAnnotation> getVariantAnnotation(@PathVariable
    @ApiParam(
        value = "Comma-separated variants (e.g., 1:g.123A>T,...)", 
        required = true, allowMultiple = true)
    List<String> variants) { /*method body*/ }
\end{minted}
% \begin{figure}[H]
%     \centering
%     \includegraphics[width=\linewidth]{figures/getvariant.png}
%     \caption*{} % prevents increment
%     % \addtocounter{figure}{-1} % restores counter
% \end{figure}
% \vspace{-2em}

In other cases, value constraints are specified in code, as seen in a DayTrader~\cite{daytrader} fragment below: the \smalltt{action} parameter supports specific values for request processing, while an invalid \smalltt{action} value triggers a 4xx or 5xx response, depending on server settings.\\
\vspace{-1.4em}
%<--CRC change
\begin{minted}{java}
public void task(HttpServletRequest req, ...)  {
    switch (req.getParameter("action")) {
    case 'q': {...} // quote
    case 'a': {...} // account
    case 'u': // update account profile
    // + 6 more cases ...}}
\end{minted}
% \begin{figure}[H]
%     \centering
%     \includegraphics[width=\linewidth]{figures/task.png}
%     \caption*{} % prevents increment
%     % \addtocounter{figure}{-1} % restores counter
% \end{figure}
% \vspace{-2em}
\tool extracts the context for an endpoint parameter, including its annotations, type, and method body, and incorporates it into an LLM prompt to extract value constraints, instructing the LLM to produce the output in a structured format illustrated with an in-context example.

\vskip 2pt
\noindent\textit{Enclosing method and class.} These represent the method and class for parameter declaration. In most frameworks, they are the endpoint class and method. For Servlets, they indicate the method and its class from which a parameter is extracted from an \smalltt{HttpServletRequest} instance, which could occur anywhere in the call chain starting at the endpoint method.

% \vskip 2pt
% \noindent\textit{Parameter annotations.} Here we store all annotations associated with an endpoint parameter.

\noindent
\textbf{Extracting inter-parameter dependencies.} API endpoints often have parameter dependencies that restrict valid request combinations. Prior work~\cite{martin:2019:catalog} identifies seven IPD types: \smalltt{AllOrNone}, \smalltt{Requires}, \smalltt{OnlyOne}, \smalltt{Or}, \smalltt{ZeroOrOne}, \smalltt{Arithmetic}, and \smalltt{Complex}. For instance, the \smalltt{OnlyOne} relation requires only one parameter, while the \smalltt{AllOrNone} relation require all or none of the parameters to be present in a valid requst. 
To extract IPDs, \tool prompts an LLM prompt with the relevant endpoint context, consisting of parameter names, types, and relevant method bodies. The prompt also includes IPD definitions and examples to teach the LLM about the relations and output formats. The LLM identifies the IPDs, determining the relation type $R$, the involved parameters $\Pi$, and the code constraints $\Gamma$, which are stored in the endpoint model (Figure~\ref{fig:endpoint-model}). 

% \paragraph{Relation type} For each relation, we store the relation type, which is one of the six IPD relations.
% \paragraph{Involved parameters in each relation} For each IPD relation, we store the parameters that are involved and it can be multiple sets for each relation.
% \paragraph{IPD constraints} Finally, we store any code constraints related to each IPD-parameter set pairs so that it can further be utilized properly for generating values correspond to endpoint parameters.

% \hl{Provide an example}

\noindent
\textbf{Extracting database operation details.}
CLDK extracts database operations based on known APIs and supported database frameworks. It maps these APIs to their corresponding CRUD operations and records the location of each call in the analysis metadata. We leverage this information to identify lines of code that contain database interactions and prioritize their coverage while testing the individual endpoints, as such operations represent an important component of the functionality provided by the endpoints.

% \paragraph{Database framework}
% \paragraph{Enclosing class name and method signature}
% \paragraph{CRUD operation type.}

\subsection{ODG Construction}
\label{subsec:odg}

% \begin{figure}
%     \centering
%     % \vspace{-8pt}
%     \includegraphics[width=0.8\linewidth]{figures/odg.pdf}
%     \caption{Operation Dependency Graph. } 
%     \label{fig:odg}
% \end{figure}

The endpoint model captures syntactic and semantic details of each endpoint without considering inter-endpoint dependencies. For instance, in PetClinic~\cite{spring-petclinic}, to add a pet, an owner ID must first be obtained via the endpoint listing all owners (\smalltt{GET} \smalltt{/owner}) or by adding a new owner (\smalltt{POST} \smalltt{/owner/\{ownerid\}}), demonstrating resource-based dependency. \tool constructs the ODG to represent such dependencies. Moreover, \tool creates a functional summary for each endpoint, used in Phase~2 to extract testing scenarios, and these summaries are linked to ODG nodes. % Figure~\ref{fig:odg} shows a sample ODG.

% \begin{figure*}[t]
%     \centering
%     \includegraphics[width=.75\linewidth]{figures/individual-endpoint.pdf}
%     \caption{The workflow for generating endpoint-focused tests.}
%     \vspace{-10pt}
%     \label{fig:individual_endpoint}
% \end{figure*}

\begin{figure*}[t]
    \centering
    \includegraphics[width=.83\linewidth]{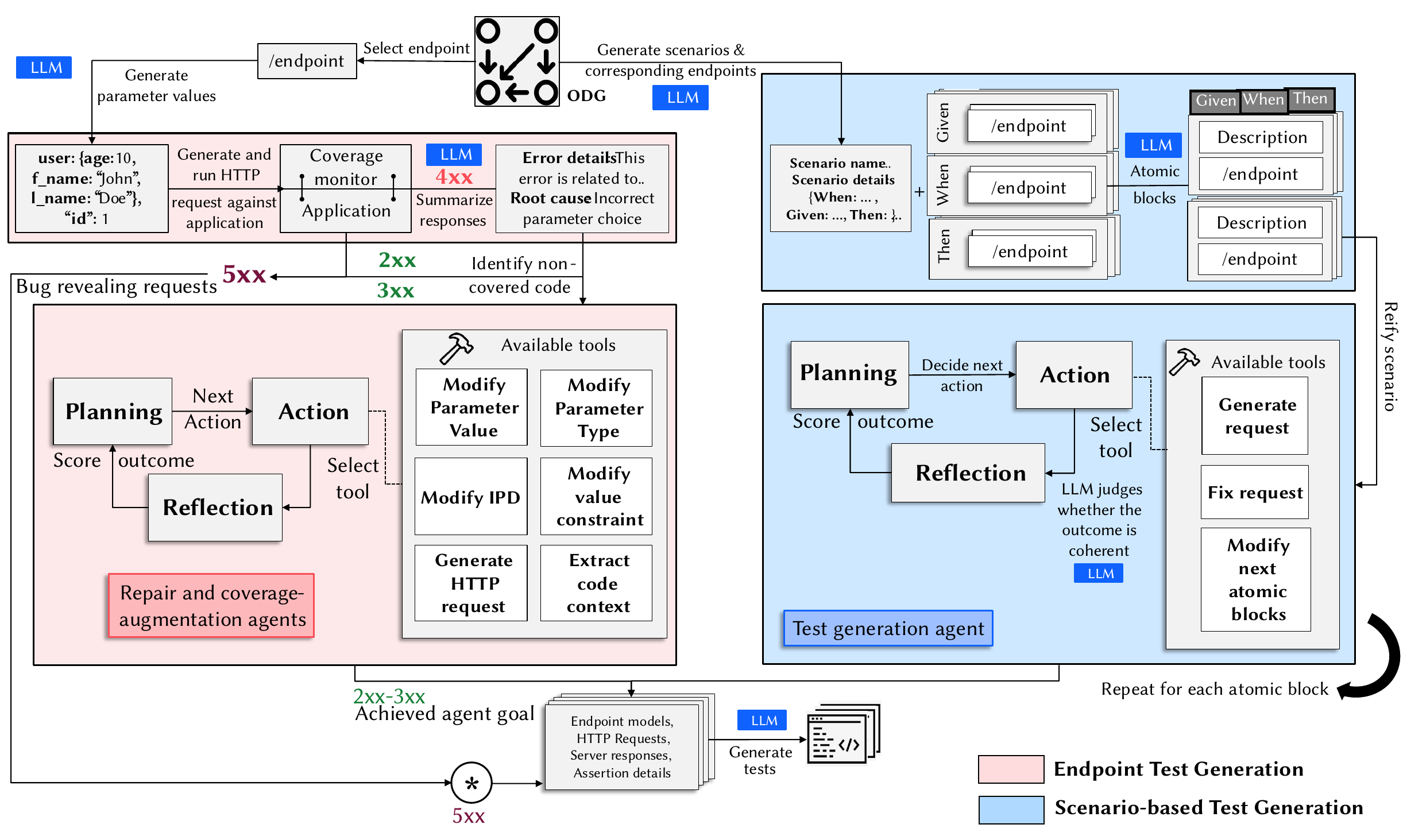}
    \vspace{-2pt}
    \caption{The workflow for generating endpoint-focused and scenario-based tests.}
    \vspace{-13pt}
    \label{fig:individual_endpoint}
\end{figure*}

Formally, the ODG $G = (V, E)$ constitutes a directed graph, wherein the nodes are representative of endpoints (or service operations) and the edges delineate the dependencies among these endpoints. Each node $v \in V$ corresponding to an endpoint encompasses the model and a comprehensive functional summary of that endpoint. An edge $(v_1, v_2) \in E$ signifies the dependency of $v_2$ on $v_1$ through one of three distinct relational types: (1) \textit{resource dependency}, where two endpoints are connected via path resources, exemplified by the PetClinic scenario; (2) \textit{producer-consumer dependency}, involving an endpoint that produces a value and another endpoint that can take that value as input; and (3) \textit{database dependency}, where one endpoint executes a write operation to a database and another endpoint retrieves data from the same database.

We use a RAFT-like approach~\cite{saha:2025:raft} to analyze path parameters and HTTP methods (\smalltt{GET}, \smalltt{POST}, \smalltt{DELETE}, etc.) for resource dependencies. For two other relations, we create prompts with endpoint and database details, and code fragments for LLMs to identify these relations. LLM calls also generate operation summaries based on provided endpoint code and other details.

\subsection{Endpoint-focused Test Generation}
\label{subsec:individual_endpoint}
Figure~\ref{fig:individual_endpoint} shows the endpoint-focused test generation workflow, which includes: (1) generating and executing HTTP requests with parameters on services, (2) fixing invalid requests, (3) enhancing code coverage, and (4) converting requests into Java tests. All steps use LLMs, with steps~2 and~3 also using agentic workflows.

\vspace{-5pt}
\subsubsection{Generation and execution of HTTP requests.}
To generate endpoint parameter values, a prompt with endpoint model details (path, HTTP method, parameter names/types, constraints, and IPDs) and related code from the call graph is created. The LLM outputs parameter-to-value mappings, forming concrete HTTP requests. These requests are executed by \tool against services monitored for coverage changes.

% First in order to extract the parameter values using LLM, we provide (a) the endpoint path, (b) the HTTP method type (e.g., GET, POST), (c) parameter names, (d) the datatype of each parameter, (e) IPDs, (f) value constraints for each parameter, and (g) relevant code snippets extracted using the call graph constructed through static analysis. 
% As output of prompting the LLM, it generates multiple sequences of parameter values focusing on covering the code as much as possible. Each sequence consists of a set of parameters and their corresponding values. Using these parameter values and other details from the endpoint model, we generate concrete HTTP requests that can be executed against a deployed application and store the response of the HTTP request. 

For each executed request, the technique checks the response code. For 4xx responses (indicating invalid requests), it invokes \tool's repair agent to fix the request. For 2xx and 3xx responses, the technique invokes the coverage-augmentation agent to increase coverage of uncovered code reachable from the endpoint method.  Both agents implement a plan-act-reflect loop consisting of a \textit{planning step} where the LLM decides on the next course of action based on available information about the task at hand, an \textit{action step} where the agent performs an action using an available set of tools, and a \textit{reflection step} where the agent ranks the outcome of the action and sends feedback for the next iteration of the loop.

% \subsubsection{Agentic approaches: Fixing HTTP requests and augmenting code coverage} In the next step, based on the response code of the executed HTTP requests, we decide whether to fixing an HTTP request or augment code coverage by generating more requests specifically focusing on the non-covered part of the code. If the response code is 4xx, which repsents that the HTTP request was ill-formed, and we attempt to fix the request by enabling \tool's agent-based HTTP request fixing pipeline. Requests with response code 2xx-3xx are further being utilized for examples while we enable \tool's coverage augmentation agent focusing on covering code that remained uncovered in the first attempt.

% \tool leverages an agentic workflow with a (a) planning phase, where LLM decides the next step based on all the information related to the problem under investigation and chooses one or more actions from the pool of actions, and (b) reflection phase, where LLM ranks the outcome and sends the feedback for the next iteration. The feedback loop will only be completed if the requirements for the task have been fulfilled or the number of turns exceeds the pre-defined limit.

\vspace{-5pt}
\subsubsection{Tools for agents.} We designed and implemented six tools that are suitable for the tasks to be performed by the repair and coverage agents (shown under ``Available tools'' in Figure~\ref{fig:individual_endpoint}).
% In this work, we have carefully selected and implemented tools that are focused to our approach, aiming to minimize the number of LLM calls compared to more generic tools used in more generic agentic workflows, e.g., SWE-Agent~\cite{sweagent}, OpenHands~\cite{openhands}. Below, we describe each of these tools in detail.

\begin{enumerate}[leftmargin=*, wide=0pt]

\item \textit{Modify parameter value.} One of the common modifications needed to fix an invalid HTTP request or increase code coverage is adjusting parameter values.
% For example, in the code below, if the \texttt{action} parameter does not match one of the accepted values, the server may return a 4xx error or fail to execute logic specific to each action type.
Selecting this action results in an LLM call with a prompt that includes relevant parameter details for the endpoint. When fixing HTTP requests, we also include the incorrect request along with an explanation generated by the LLM while selecting an action. Additionally, if the LLM determines that more code context is necessary, we include it in the prompt. For the coverage-augmentation, we supply the LLM with uncovered lines. % ensuring necessary adjustments to either correct the request or extend code coverage.
    % \begin{minted}[escapeinside=@@, frame=lines,framesep=1mm,baselinestretch=0.5, fontsize=\scriptsize, breaklines, breakanywhere, linenos,numbersep=2pt, highlightlines={}]{java}
    % public void performTask(HttpServletRequest req, HttpServletResponse resp){...
    %         String action = req.getParameter("action");
    %         switch (action) {
    %             case 'q': {...} // quote
    %             case 'a': {...} // account
    %             case 'u': {...} // update account profile
    %             ...
    % \end{minted}

\item \textit{Modify parameter type.} This action is designed for handling application frameworks (e.g., Servlet) for which we obtain parameter types via LLM calls. If during the planning step, the agent reasons that the assigned type of a parameter causes an invalid request or uncovered lines, it can rectify that mistake via this action. The output of this action consists of new requests with parameter values generated in accordance with modified parameter types.
% are formed after changing the datatype of the selected parameters.

\item \textit{Modify IPD.} Similar to the parameter-type-update action, this action updates an IPD that was initially obtained via LLM prompting. Based on the new IPD, a new request is formed.

\item \textit{Update value constraint.} This action updates a previously extracted value constraint for a parameter, and forms a new HTTP request based on that.

\item \textit{Generate requests.} With this action, the agent generates more requests for the same endpoint or another endpoint, whose invocation may be a prerequisite for fixing an invalid request or covering more code in the endpoint under consideration.
% For instance, in the previous code snippet, \tool requires to create a valid register and login with the correct username and password. Another use case is processing other endpoint as a pre-requisite, which can both help fixing HTTP request and covering more code in the endpoint under test.

\item \textit{Extract additional code context.} Often fixing an HTTP request or covering additional code requires code-related details that may not be available in the initial prompt. With this action, the agent can request more code context by selecting a CLDK API~\cite{cldk} to be invoked (from a list of APIs provided to the agent). For instance, the agent can request information about callees of an endpoint method.

% For instance, in the below code, to understand how logic when \texttt{action} parameter is set to \texttt{updateConfig}, one requires to look into the details of the callees, which is case is the \texttt{doConfigUpdate(HttpServletRequest, HttpServletResponse)} method. If at point of time, LLM understands that it requires more code context, it can call CLDK APIs~\cite{} through function calling and gather additional code context.

% \begin{minted}[escapeinside=@@, frame=lines,framesep=1mm,baselinestretch=0.5, fontsize=\scriptsize, breaklines, breakanywhere, linenos,numbersep=2pt, highlightlines={7, 9, 10}]{java}
% @Override
% public void service(HttpServletRequest req, HttpServletResponse resp) throws ServletException, IOException {
% ...
% try {
%   action = req.getParameter("action");
%   if (action == null) {
%     doConfigDisplay(req, resp, result + "<b><br>Current DayTrader Configuration:</br></b>");
%     return;
%   } else if (action.equals("updateConfig")) {
%     doConfigUpdate(req, resp);
%   } ...
% }
% void doConfigUpdate(HttpServletRequest req, HttpServletResponse resp) throws Exception {...
%     String orderProcessingModeStr = req.getParameter("OrderProcessingMode");
%     if (orderProcessingModeStr != null) {
%       try {
%         int i = Integer.parseInt(orderProcessingModeStr);
%         if ((i >= 0) && (i < TradeConfig.getOrderProcessingModeNames().length)) ...
%       } catch (Exception e) {...} ...} 
% \end{minted}

\end{enumerate}

\vspace{-5pt}
\subsubsection{Repair agent.} The repair agent attempts to fix invalid HTTP requests. During the planning step, the agent is presented with the request details, along with LLM-generated problem summary of the error response from the server. As the raw server response can be overly verbose, making it hard to pinpoint the issue, we use an LLM to summarize the response, which makes the agent's planning step easier. Based on the presented information, the agent selects the next actions to execute (we limit this to two actions to control the computational cost). Along with the actions, the agent also generates the rationale for its decision. After executing a selected action, the agent reflects on the outcome by comparing the summarized server responses before and after the action to determine whether the action addresses the problem with the original invalid request. For instance, upon receiving a 4xx response and the corresponding server message, the repair agent selects suitable tools to regenerate the request and re-evaluates the response. This process continues until a 200 status code is obtained or a predefined upper bound on the number of attempts is reached. A scoring mechanism guides the agent’s decisions at each iteration when the goal remains unmet.

\vspace{-5pt}
\subsubsection{Coverage-augmentation agent.} This agent is tasked with generated HTTP requests targeted at covering specific (uncovered) lines of code. During the planning step, the agent is provided with the uncovered reachable lines of code and other relevant information about the tested endpoint. The agent selects up to two actions to execute next, along with the reasoning behind its choices. After performing the action, which results in execution of newly generated HTTP requests against the endpoint, the agent reflects on the outcome by comparing the previously uncovered lines with the newly covered lines after the execution of the new requests.

All the tools (or actions), except the code-context action, generate one or more HTTP requests. The code-context action produces code-related details returned from the CLDK API chosen. During reflection for this action, the agent evaluates whether it contributes meaningfully to facilitating future request generation in subsequent iterations. As part of reflection, the agent computes a score in the range [0, 1], indicating the effectiveness of the action, and creates a comment explaining the rationale for the score. These are used in the next iteration to guide the selection of the next action.

\vspace{-5pt}
\subsubsection{Test generation step}
In this step, we convert selected requests into executable test cases. Specifically, we focus on two types of requests: (1) those that contribute to code coverage and (2) those that trigger server failures (i.e., 5xx response codes). For identifying coverage-contributing requests, we rely on a coverage monitoring agent deployed with the application. Once these requests are identified, we extract endpoint information and generate corresponding test cases using a code skeleton defined through Jinja templates~\cite{supplementary}.
% The generated tests include appropriate package declarations and are organized according to the application's source code directory structure. This design choice aids developers in easily mapping each test to its corresponding endpoint class, improving traceability and maintainability.
\vspace{-5pt}
\subsection{Scenario-based Test Generation}

Scenario-based test generation focuses on creating meaningful sequences of API calls for exercising application use cases, as illustrated by the PetClinic test case in Listing~\ref{lst:motivating-example-2}.
% While generating tests for individual endpoints helps check the functionality of an endpoint in isolation but real-life application use cases involve more complex objectives. For instance, let's go back to the \texttt{Spring-PetClinic} example. One application use case can be ``\textit{viewing the list of veterinaries and add a visit to a selected veterinary by providing details regarding an existing pet and its parent}''. In this use case, based on the available endpoints, a series of endpoints need to be exercised in a particular sequence.
The right side of Figure~\ref{fig:individual_endpoint} illustrates the workflow for generating scenario-based tests, which consists of four steps. In the first step, \tool extract test scenarios from the application code and map each scenario to a sequence of endpoints via an LLM call. The second step decomposes a scenario into a sequence of atomic blocks, where each atomic block achieves a specific step of a test scenario and is associated with one endpoint. In the third step, \tool employs an agentic approach to generate the test fragment for each atomic block, using the generated information for a block to process subsequent blocks. The final step composes the test fragments together to create an executable JUnit test case for the scenario.

% To do that, first, we extract various scenarios using LLM. Then, we map these scenarios to concrete endpoints. Next, we break each scenario into a series of atomic blocks, where each atomic block achieves a specific sub-task of an application scenario and can only be associated with one endpoint. Then, we use an agentic approach to generate a test by processing one atomic block at a time and using the processed information for the subsequent atomic blocks. Finally, we convert these tests into an \texttt{JUnit} tests using \texttt{Rest-Assured} framework. In Figure~\ref{fig:scenario}, we depict each phase of scenario-driven test generation.

% \begin{figure}[t]
%     \centering
%     % \vspace{-8pt}
%     \includegraphics[width=\columnwidth]{figures/scenario-agent.pdf}
%     \caption{The workflow for scenario-based test generation.} 
%     \label{fig:scenario}
% \end{figure}

\vspace{-5pt}
\subsubsection{Generation of test scenarios and related endpoints.}
To generate test scenarios, our approach constructs an LLM prompt that includes endpoint information and functional summaries. The prompt instructs the model to produce scenarios aligned with business use cases using Gherkin-like syntax~\cite{gherkin:syntax}. Each scenario follows a structured format with a scenario name, a \texttt{given} clause (preconditions), a \texttt{when} clause (actions), and a \texttt{then} clause (expected outcomes). We experimented with various formats and found the Gherkin-like style most effective, even enabling smaller models to generate coherent, meaningful scenarios. Figure~\ref{fig:scenario-example} shows an example generated for the PetClinic application~\cite{spring-petclinic}.

\begin{figure}[t]
\vspace{-1em}
% \sciptsize
% <--CRC Change
\begin{minted}[style=friendly, fontsize=\scriptsize]{yaml}
scenario: View list of veterinaries and visit a selected veterinary
given:
  - A vet with ID 1 exists in the system
  - An owner with ID 1 exists in the system having a pet with ID 1
when:
  - A request is made to the 'showVetList' endpoint with vet ID 1
  - The vet's details are viewed
  - A request is made to 'initNewVisitForm' to initialize a new
    visit form
  - The form is filled and submitted to 'processNewVisitForm'
then:
  - The vet's information is displayed
  - The new visit is added to the pet's record in the database
\end{minted}
% \includegraphics[width=\linewidth]{figures/scenario.png}
% \vspace{-1.4em}
\caption{Sample test scenario extracted by \tool.} 
\label{fig:scenario-example}
\end{figure}
% \begin{minted}

% \vspace{-5pt}
\subsubsection{Decomposing scenarios into atomic blocks.} Each clause of a scenario can be associated with one or more endpoints. If a \smalltt{given}, \smalltt{when}, or \smalltt{then} clause has more than one endpoint, \tool decomposes it into more granular tasks, or \textit{atomic block}, so that processing a single endpoint can help achieve that task. The prompt instructs the LLM to divide a scenario into atomic blocks and provide details on how the blocks are related. For the example scenario in Figure~\ref{fig:scenario-example}, the \smalltt{when} can be divided into more granular tasks of retrieving vet information, initializing a new visit form, and creating a new visit, where the last two tasks share the same pet and owner ID. The information about decomposed blocks is then fed to the test-generation agent to create concrete HTTP requests and test case. 

\vspace{-5pt}
\subsubsection{Test-generation agent.} The agent processes each atomic block of a scenario to generate concrete requests, and uses the responses from those requests to process the subsequent blocks. In the planning step, the agents decide on the next course of action and selects from a set of available tools. We designed three tools tailored to the task of converting a sequence of atomic blocks for a scenario to a sequence of test fragments.

% \paragraph{Available tools} The selected tools are very specialized for this purpose. We have three tools--(a) generating HTTP requests with all the available information on the atomic block under process and the outcome of the previous blocks, (b) fixing HTTP requests if it does not coherent with the description of the block, and (c) modifying the task description of the subsequent blocks based on the outcome of the block under process.
\begin{enumerate}[leftmargin=*, wide=0pt]

\item \textit{Generate requests.} With this tool, the agent generates HTTP requests for the endpoint corresponding to an atomic block. The relevant context in the LLM prompt includes endpoint details, the scenario description, the task for the particular block, the outcomes of preceding blocks (requests and corresponding responses), and in-content examples. The generated requests are executed against the deployed application to obtain the responses. 

    % \item \textit{LLM as a judge} Here, we prompt LLM with all the generated requests and responses to identify the requests that are in accordance to the task description. If it identifies that none of the requests is related to the task, then it provides a justification of why it thinks that and we utilize that in the request fixing tool.

\item \textit{Fix request.} With this tool, the agent attempts to modify the parameter values of a previously generated request so that the request aligns with the task description for an atomic block.
% If at any point in time, the agent finds that none of the generated request conforms to the task, then with the justification of that choice, we provide all the details for the atomic block under process and ask LLM to modify the parameter values to align with the task description.

\item \textit{Modify subsequent atomic blocks.} The outcome of one block can require modifications to subsequent blocks. Consider again the scenario in Figure~\ref{fig:scenario-example}. Suppose that while processing the task for the \smalltt{given} clause, the agent finds that there exists a vet with ID~2; this ID can also be used for the scenario, but it requires the descriptions of the subsequent blocks to be updated. The agent uses the scenario description, the current block being processed, and its requests and  outcomes to modify descriptions of the subsequent blocks (if required) such that the overall goal of the scenario is preserved while minor details (e.g., vet ID) are updated.

\end{enumerate}

% \paragraph{Planning.} The planning phase is more deterministic than our other agents. For processing a block, we always start with generating the concrete requests. Then we apply the LLM as a judge tool to let us know if the outcome of the requests matches with the description of the block. If it does, we ask LLM if the outcome requires any modification to the subsequent blocks. If it does, we apply the tool to modify the subsequent blocks. If the LLM does not accept the outcome, then we apply the fixing step. We only get out of the loop when LLM accepts the outcome of the request. We apply this step for each atomic block in the scenario.

% \paragraph{Reflection.}
After executing an action, the agent reflects on the outcome of the executed requests  to determine whether it aligns with the description of the current block under processing. 
% Here, we prompt LLM with all the generated requests and responses to identify the requests that are in accordance to the task description.
If it determines that the requests are unrelated to the block, it provides a justification for its decision. If the outcome aligns, the agent also determines whether any subsequent blocks in the scenario need modifications. The information from this step then feeds into the next iteration of planning to select action to be performed.

%and we utilize that in the request fixing tool.
% Here, we use LLM as a judge to deterministically define the next step to take.

\vspace{-5pt}
\subsubsection{Generate Tests.} After the test-generation agent completes, \tool collects all relevant information—including the scenario description, the atomic block sequences, the HTTP requests, and the responses—and prompts an LLM to generate a test using the Rest-Assured framework~\cite{restassured}. Given the structured nature of this data and the simplicity of Rest-Assured’s syntax, the LLM consistently produces compilable tests. Finally, \tool adds the scenario description and Java package information to the final test.

% % \subsection{Prompts}

% \begin{figure*}
%     \centering
%     % \vspace{-8pt}
%     \includegraphics[width=\linewidth]{figures/short_prompt.pdf}
%     \caption{SAINT prompts. The \colorbox{cyan!20}{cyan box} and \colorbox{yellow!20}{yellow box} represents the required parameters and optional parameters, respectively. TU: Tool usage.} 
%     \label{fig:prompt}
% \end{figure*}
% \input{prompts}
\section{Evaluation}
\label{sec:experiment}

% In this section, we discuss our evaluation, the experiment setup, dataset characteristics and the research questions that we attempted to answer. From the research question point, we attempted to answer five questions, starting from measuring effectiveness of \tool is in terms of (a) generating test for individual endpoint and (b) a sequence of endpoints guided by application use-cases. We also conducted a survey of \hl{x} developers to understand the quality of the extracted scenarios and the generated tests surrounding them. Moreover, we looked into the fault detection capability of \tool in terms of generating tests for individual endpoints and scenarios. Finally, we performed an ablation study to understand the effectiveness of various parts of \tool. Below, we list all these research questions.

Our evaluation focuses on the following five research questions:

\begin{itemize}[leftmargin=*]
    \item \textbf{RQ1}: (Coverage) How does \tool compare with EvoMaster~\cite{arcuri_evomaster_2018} in terms of code coverage and database interaction coverage?
    \item \textbf{RQ2:} (Scenario Effectiveness) How effective is \tool in generating scenario-based tests?
    \item \textbf{RQ3:} (Developer survey) How do developers perceive the scenario-based tests generated by \tool in terms of their usefulness?
    \item \textbf{RQ4:} (Fault Triggering) How does \tool compare with EvoMaster in terms of server failures triggered?
    \item \textbf{RQ5:} (Ablation) How do ODG construction, IPDs and value constraints extraction, repair agent, and coverage-augmentation agent contribute to \tool's effectiveness in code coverage?
    % \item \textbf{RQ6:} \hl{keep this RQ?} How does \tool compare with the state-of-the-art REST API specification extraction tool Respector~\cite{huang:2024:respector} in terms of endpoint signature and constraint extraction?
    % \item \textbf{RQ7:} \hl{TBD} Application of generated tests for differential testing in an app modernization scenario
\end{itemize}

% For evaluating RQ3, we have to assess (via manual inspection) whether (1) an LLM-created test scenario is valid for the app under test, and (2) the concrete test sequence generated for a test scenario actually implements the scenario fully. In terms of automated metrics, we can measure additional coverage gain from scenario-based tests over the endpoint-focused tests and EvoMaster-generated tests. We can evaluate valid scenarios/sequences based on whether they could be created by exploring producer-consumer relations, CRUD-based ordering, resource-based dependencies, or operation prerequisites (as existing tools do) or whether such exploration is insufficient for generating the tests.

% \hl{RQ on effectiveness of IPD and value constraint extraction?}

\subsection{Experiment Setup}
\label{sec:rqs}

We evaluated our approach on two types of applications: (1) REST APIs with OpenAPI specifications and (2) enterprise Java applications without OpenAPI specifications. The first group includes four APIs from the EvoMaster benchmark—Feature-Service, Genome Nexus, LanguageTool, and RestCountries. The second group includes three open-source Java applications (DayTrader, JPetStore, and PetClinic) and one proprietary enterprise application (App-X). These open-source applications have been used in prior empirical evaluations~\cite{pan2025aster, nitin2022cargo}.
% We added three more applications to the dataset after excluding those that could not be deployed or crashed frequently.
Table~\ref{tb:dataset} summarizes the dataset.

\begin{table}[t]
\caption{Java applications used in the evaluation.}
\centering
\renewcommand{\arraystretch}{1.0}
\resizebox{0.9\linewidth}{!}{
\begin{tabular}{lcccccc}
\toprule
\shortstack{\textbf{Dataset}} & 
\shortstack{\textbf{Framework}} & 
\shortstack{\textbf{Java}\\\textbf{version} } & 
\shortstack{\textbf{OpenAPI} \\ \textbf{spec?}} & 
\shortstack{\textbf{NCLOC}} & 
\shortstack{\textbf{\# of} \\ \textbf{classes}} & 
\shortstack{\textbf{\# of} \\ \textbf{endpoints}}\\
\midrule
DayTrader & Servlet & 8 & \textcolor{red}{\textbf{X}} & 11409 & 141 & 113\\
PetClinic & Spring & 17 & \textcolor{red}{\textbf{X}} & 790 & 24 & 17\\
JPetStore & Stripes & 8 & \textcolor{red}{\textbf{X}} & 1409 & 24 & 21\\
Restcountries & Jax-rs & 8 & \textcolor{teal}{\textbf{\checkmark}} & 1619 & 23 & 27\\
Feature-service & Jax-rs & 8 & \textcolor{teal}{\textbf{\checkmark}} & 1688 & 21 & 18\\
Genome-Nexus & Spring & 8 & \textcolor{teal}{\textbf{\checkmark}} & 22143  & 74 & 48 \\
Languagetool & HttpServer & 8 & \textcolor{teal}{\textbf{\checkmark}} & 113170 & 37 & 6\\
App-X & Servlet & 11 & \textcolor{red}{\textbf{X}} & 1255 & 24 &23\\
\bottomrule
\end{tabular}}
\label{tb:dataset}
\vspace{2pt}
\end{table}

% Acme-air
% 6. A couple of services from the Respector paper benchmark: check which ones can be deployed https://dl.acm.org/doi/pdf/10.1145/3597503.3639137
\vskip 2pt
\noindent\textit{Testing Tools.} We compared \tool with EvoMaster, a state-of-the-art white-box test-generation tool~\cite{arcuri_evomaster_2018}. Although EvoMaster uses white-box techniques, it still requires an OpenAPI specification, so our comparison with it is limited to the four applications with specifications. We tried to generate OpenAPI specifications for the other applications using popular specification-generation tools (e.g., springdoc-openapi~\cite{springdoc}, SpringFox~\cite{springfox}), but they produced incomplete specifications that would not work with EvoMaster without extensive manual effort. %; so we excluded those applications.
EvoMaster was run on each application for one hour. % (using a random seed).
Because EvoMaster’s JVM agent conflicts with the JaCoCo agent~\cite{arcuri_survey}, we used EvoMaster to generate tests and then executed the tests with JaCoCo to obtain coverage information.

\vskip 2pt
\noindent\textit{LLMs.} We selected the models based on size, cost, family, and popularity---categorized as small (IBM Granite 3.1–8B, Meta Llama 3.1–8B), medium (Devstral-24B, DeepSeek-R1-distill-Qwen-32B), and large (GPT-o1). Given the high computational cost of cross-dataset evaluation, each model was executed twice using parameters from prior work~\cite{pan2025aster}, with the temperature set to 0.2 to produce stable yet diverse outputs.
% Model details are summarized in Table~\ref{tb:model}
\tool uses 25 unique prompts in its pipeline, which are available in our artifact~\cite{supplementary}.

\begin{figure*}[t]
    \centering
    % \vspace{-8pt}
    % \includegraphics[width=\linewidth]{figures/rq1-wide.pdf}
    \includegraphics[width=\linewidth]{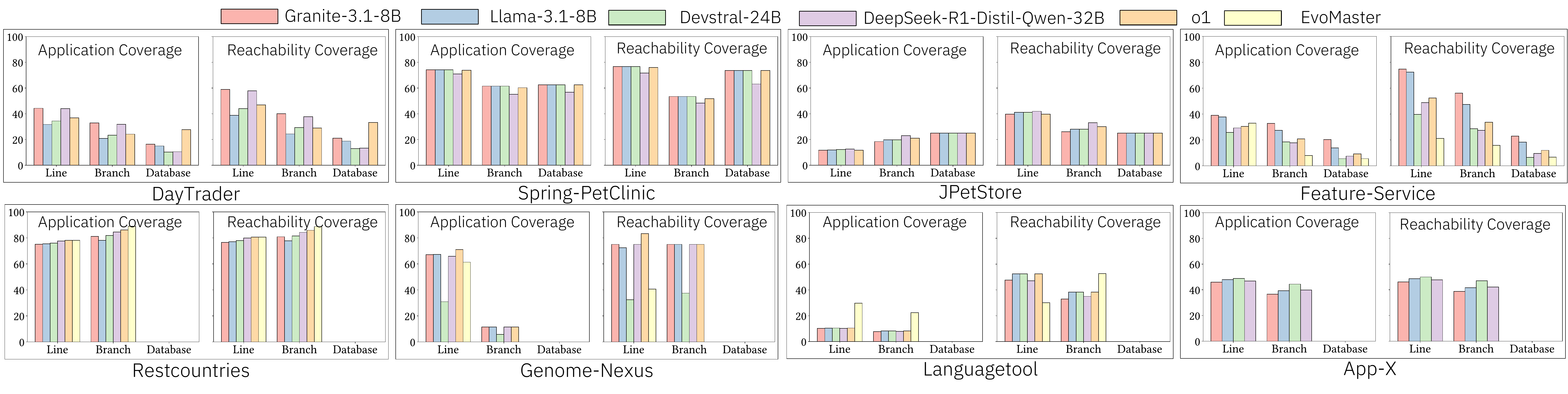}
    % \vspace{-5pt}
    \caption{Application and reachability coverage of lines, branches, and database interaction points.}
    \vspace{-12pt}
    \label{fig:rq1}
\end{figure*}

\vskip 2pt
\noindent\textit{Metrics.} We used line, branch, and database line coverage as evaluation metrics. %, using CLDK~\cite{cldk} to identify database interaction points.
Using the JaCoCo agent~\cite{jacoco}, we collected real-time coverage during execution. We also computed \emph{reachability coverage} by analyzing call chains starting from each endpoint and identifying the reachable methods; this lets us measure coverage within the effective execution scope---the portion of the code that is reachable from the endpoints.
% Additionally, we measured operation coverage as the proportion of API operations exercised.
Scenario quality metrics included scenario length, count, endpoints exercised, and other structural features.

\subsection{Experiment Results}

\subsubsection{RQ1: Code coverage}
% In this RQ, we evaluate \tool's capability to generate endpoint-focused tests against two sets of applications: those with and without OpenAPI specifications. 
Figure~\ref{fig:rq1} presents the results on code coverage achieved by the endpoint-focused tests generated by \tool with different model configurations, and compares the results with coverage achieved by EvoMaster.

\vskip 2pt
\noindent\textit{Applications without OpenAPI Specifications.}
This application category consists of PetClinic, DayTrader, JPetStore, and App-X. Among these, \tool could not measure database coverage for App-X, which uses DB2—a database currently not supported by CLDK. In terms of database line coverage, GPT-o1 performed the best overall. However, for PetClinic and JPetStore, all models achieved comparable coverage. A manual inspection of JPetStore identified four database call sites, three of which involve complex conditional logic—posing challenges for all models. Interestingly, on PetClinic, all models achieved similar line, branch, and database coverage. Further analysis revealed that gaining additional coverage would require solving intricate constraints, which \tool currently does not support. Overall, the performance of different models is comparable, with the smaller Granite-8B often matching or outperforming larger models like o1. The generated tests are implemented in Java using the Rest-assured framework~\cite{restassured}. Two sample tests are shown below:
%<--CRC change
\begin{minted}[fontsize=\scriptsize]{java}
@Test public void testProcessUpdateForm1() {
    Map<String, String> pet = new HashMap<>();
    pet.put("name", "Fido");
    pet.put("birthDate", "2010-05-01");
    given().queryParam("pet", pet)
        .pathParam("ownerId", "1")
        .pathParam("petId", "1").when()   .post("http://localhost:8080/owners/{ownerId}/pets/{petId}/edit")
    .then().statusCode(302); }
@Test public void testProcessUpdateForm2() {
    Map<String, String> pet = new HashMap<>();
    pet.put("name", "Buddy");
    pet.put("birthDate", "2012-08-15");
    given().queryParam("pet", pet)
        .pathParam("ownerId", "10")
        .pathParam("petId", "2").when()
    .post("http://localhost:8080/owners/{ownerId}/pets/{petId}/edit")
    .then().statusCode(500); } ...}
\end{minted}

% \begin{figure}[H]
%     \centering
%     \includegraphics[width=\linewidth]{figures/testprocessupdateform.png}
%     \caption*{} % prevents increment
%     % \addtocounter{figure}{-1} % restores counter
% \end{figure}

The tests simulate a user updating the details of an existing pet. \tool generates multiple tests covering both positive and negative paths. In the first scenario, it uses valid owner and pet IDs to update pet information, resulting in a successful 200 response. In contrast, the second scenario uses invalid (non-existent) IDs, which leads to a 500 server error. All tests generated by \tool are compilable and executable, requiring no edits.

\begin{findingbox}{1}
{\tool can generate tests for applications without OpenAPI specifications with high code coverage (10\%--80\%). Also, with \tool, smaller models (8B parameters) can achieve similar or better coverage compared to bigger models such as GPT-o1.}
\end{findingbox}

\vskip 2pt
\noindent\textit{Applications with OpenAPI Specification.}
Among all the applications in this category, Restcountries and Languagetool do not interact with any database, while Genome-Nexus uses MongoDB, which is currently unsupported by CLDK. Consequently, database coverage is only reported for Feature-service. In this case, \tool achieves notably higher database coverage, successfully exercising multiple database interaction points. We also found that \tool can identify more endpoints than those defined in the OpenAPI specification—most notably in Genome-Nexus, where the specification lists 23 endpoints, but \tool detected 48.

In terms of line and branch reachability coverage, \tool outperforms EvoMaster for Feature-service and Genome-Nexus (+50.5\% and +22.0\% in line coverage). For Restcountries, \tool's performance is slightly lower (-0.9\%) in branch coverage). For Languagetool, \tool's achieves considerably lower (-19.3\%) application coverage than EvoMaster. However, when compared on reachability coverage, they are comparable. The discrepancies observed in LanguageTool arise from an incomplete API specification: the OpenAPI specification lists only two endpoints, while more endpoints exist in the implementation. This is not uncommon, as OpenAPI specifications are primarily written for external users and often omit endpoints meant for internal usage. 
\begin{findingbox}{2}
{Compared to EvoMaster, \tool achieves similar or considerably better coverage, with the coverage difference ranging from -0.9\% to +50.5\%.}
\end{findingbox}

% \rangeet{In terms of operation coverage, in most cases, except for Genome-Nexus with o1 model (33.3\%), the operation coverage for all the applications under test in very high (mean $\sim$ 100\%). }

\subsubsection{RQ2: Effectiveness of scenario-based test generation}
\begin{table}[t]
  \centering
  \setlength{\tabcolsep}{2pt} 
  \caption{Effectiveness of scenarios and scenario-based tests.}
   \resizebox{1\linewidth}{!}{  \begin{tabular}{c|c|c|c|c|c|c|c|c|c}
   \toprule
    \multirow{2}[0]{*}{App} & \multirow{2}[0]{*}{Models} & \multirow{2}[0]{*}{\shortstack{\# of \\ scenarios}} & \multirow{2}[0]{*}{\shortstack{Sequence \\ length}} & \multirow{2}[0]{*}{\shortstack{Scenarios w/ \\ >1 class (\%)}} & \multicolumn{4}{c|}{Coverage (\%)} &  \multirow{2}[0]{*}{\shortstack{(good, bad) \\ path  scenario (\%)}} \\
    \cline{6-9}
          &       &       &   &    & \multicolumn{1}{l|}{Line} & \multicolumn{1}{l|}{Branch} & \multicolumn{1}{l|}{DB} &   \multicolumn{1}{l|}{Operation}    &         \\
    \midrule
    \multirow{5}[0]{*}{PC} & G-8B    & 4.0   & 2.8   & 87.5  & 57.5  & 41.7  & 47.5  & 52.9  & (87.5, 12.5)         \\
    % \cline{2-10}
          & L-8B  &  7.0   & 2.4   & 14.3  & 64.3  & 53.2  & 47.9  & 61.8  & (71.4, 28.6)         \\
    
    % \cline{2-10}
          & DV-24B    &       7.0   & 2.3   & 5.6   & 68.4  & 48.1  & 50.5  & 82.4  & (78.9, 21.1)      \\
    % \cline{2-10}
          & DS-32B    &      4.5   & 2.4   & 32.5  & 61.6  & 50.6  & 55.5  & 52.9  & (90.0, 10.0) \\
    % \cline{2-10}
          & o1    &       6.0   & 3.2   & 50.0  & 61.2  & 37.2  & 42.1  & 85.3  & (66.7, 33.3)       \\
    \midrule
    \multirow{5}[0]{*}{DT} & G-8B    &  6.5  & 1.5  &  18.1     &  7.8     &  6.4     &   7.4    &    23.2   &     (100.0, 0.0)    \\
    % \cline{2-10}
          & L-8B    &  8     &   2.4    &   60.0  &      8.1   &   6.9    &  5.6     &  25.6  &  (90.0, 10.0)   \\
    
    % \cline{2-10}
          & DV-24B    &   18    &     2.1  &  25.3   & 8.0 &    6.4   &    7.4   &   43.9    &     (68.2, 31.8)    \\
    % \cline{2-10}
          & DS-32B    & 14      &     2.0  & 13.9   & 8.1  &  6.4     &   7.4    &    42.7   &   (92.2, 7.8)      \\
    % \cline{2-10}
          & o1    &   8    &   4.2    &   62.5   &10.2 &   10.1    &    5.6   &  80.5     &   (93.7, 6.3)     \\
    \midrule
    \multirow{5}[0]{*}{JP} & G-8B    &  4.5  & 4 &  90.0 &  7.8     &   5.3    &     -  &   45.2    &   (35.0, 65.0)         \\
    % \cline{2-10}
          & L-8B    &   10    &    2.5   &  31.2   & 8.0 &    5.3   &  -     &  69.0     &      (43.8, 56.2)   \\
    
    % \cline{2-10}
          & DV-24B    &   10    &     4.4  &   65.0  & 8.1 &   5.3    &   -    &   97.2    &   (25.0, 75.0)      \\
    % \cline{2-10}
          & DS-32B    &   7.5    &   3.2    &  42.0  & 8.1  &    5.3   &    -   &  85.7     &     (45.5, 54.5)    \\
    % \cline{2-10}
          & o1    &    5   &     5.2  &    20.0  & 8.1&   5.3    &    -   &    100   &   (60.0, 40.0)     \\
    \midrule
    \multirow{5}[0]{*}{FS} & G-8B    &    5.5   & 2.6   & 60.7  & 21.2  & 11.3  & 0.0   & 61.1  & (21.4, 78.6)     \\
    % \cline{2-10}
          & L-8B    &      14.5  & 2.3   & 46.1  & 26.6  & 16.1  & 3.7   & 80.6  & (28.4, 71.6)         \\
    
    % \cline{2-10}
          & DV-24B    &       11.5  & 2.0   & 31.5  & 19.0  & 9.5   & 1.5   & 83.3  & (20.4, 79.6)   \\
    % \cline{2-10}
          & DS-32B    &       4.0   & 5.9   & 70.0  & 13.4  & 8.3   & 5.9   & 77.8  & (36.7, 63.3)   \\
    % \cline{2-10}
          & o1    &       3.5   & 4.9   & 70.0  & 36.5  & 33.9  & 19.8  & 75.0  & (65.0, 35.0)   \\
    \midrule
    \multirow{5}[0]{*}{RC} & G-8B    &   5.0   & 2.1   & 10.0  & 38.7  & 30.0  & - & 33.9  & (70.0, 30.0)        \\
    % \cline{2-10}
          & L-8B    &       11.0  & 3.0   & 20.0  & 30.0  & 20.2  & - & 26.8  & (29.2, 70.8)        \\
    
    % \cline{2-10}
          & DV-24B    &       15.5  & 1.5   & 12.5  & 63.8  & 63.6  & - & 51.8  & (61.2, 38.8)        \\
    % \cline{2-10}
          & DS-32B    &    10.5   &   1.4    &  16.7  &  44.9 &   43.6 & -    &    42.8   &      (59.7, 40.3)        \\
    % \cline{2-10}
          & o1    &       7.5   & 3.9   & 100.0 & 65.4  & 59.3  & - & 94.6  & (80.0, 20.0) \\
    \midrule
    \multirow{5}[0]{*}{GN} & G-8B    &    7.5   & 1.9   & 81.3  & 49.0  & 1.9   & - & 19.8  & (72.3, 27.7)         \\
    % \cline{2-10}
          & L-8B    &   13.0    &    1.7   &   0.0  &    54.2   &    1.9      &    - & 45.83 & (69.3, 30.8)     \\
    
    % \cline{2-10}
          & DV-24B    &       14.0  & 1.8   & 36.4  & 53.1  & 7.7   & - & 41.7  & (74.6, 25.4)       \\
    % \cline{2-10}
          & DS-32B    &       6.0   & 2.3   & 95.0  & 49.1  & 1.9   & - & 19.8  & (100.0, 0.0)      \\
    % \cline{2-10}
          & o1    &       8.0   & 2.5   & 25.0  & 54.5  & 7.7   & - & 39.6  & (75.0, 25.0)       \\
    \midrule
    % \multirow{5}[0]{*}{LT} & G-8B    &    &   &       &       &       &       &       &         \\
    % \cline{2-10}
    %       & L-8B    &       &       &     &  &       &       &       &         \\
    % \cline{2-10}
    %       & DS-32B    &       &       &    &   &       &       &       &         \\
    % \cline{2-10}
    %       & DV-24B    &       &       &     &  &       &       &       &         \\
    % \cline{2-10}
    %       & o1    &       &       &      & &       &       &       &        \\
    % \hline
    \multirow{5}[0]{*}{A-X} & G-8B    &    8.0   & 2.2   & 48.4  & 24.1  & 11.6  & - & 28.3  & (56.3, 43.7)       \\
    % \cline{2-10}
          & L-8B    &      16.5  & 2.7   & 73.3  & 27.9  & 14.4  & - & 56.5  & (61.7, 38.3)      \\
    
    % \cline{2-10}
          & DV-24B    &      10.0  & 2.2   & 35.0  & 30.9  & 18.1  & - & 50.0  & (65.0, 35.0)     \\
    % \cline{2-10}
          & DS-32B    &      4.0   & 3.1   & 80.0  & 26.0  & 13.9  & - & 32.6  & (90.0, 10.0)      \\
    \bottomrule
    % \multirow{5}[0]{*}{A-Y} & G-8B    &    &   &       &       &       &       &       &         \\
    % \cline{2-10}
    %       & L-8B    &       &       &     &  &       &       &       &         \\
    % \cline{2-10}
    %       & DS-32B    &       &       &    &   &       &       &       &         \\
    % \cline{2-10}
    %       & DV-24B    &       &       &     &  &       &       &       &         \\

    %     \hline
    \end{tabular}}\\
    \scriptsize * G-8B: Granite-3.1.-8B, L-8B: Llama-3.1.-8B, DS-32B: DeepSeek-R1-Distil-Qwen-32B, DV-24B: Devstral Small, PC:  PetClinic, DT: DayTrader, JP: JPetStore, FS: Feature-Service, RC: Restcountries, GN: Genome-Nexus, A-X: App-X
  \label{tb:scenario}%
  \vspace{3pt}
\end{table}

Besides testing individual endpoints, a key feature of \tool is extracting test scenarios and converting them into Java tests.
% First, \tool employs an LLM to generate a brief summary of each endpoint's behavior. We use the ODG to extract and convert application use cases into Java tests.
To assess the quality of extracted test scenarios, we evaluate the number, length, endpoint class diversity, and good-path versus bad-path distribution of scenarios, as shown in Table~\ref{tb:scenario}. We classify scenarios as good-path if they result in a 2xx status code, and as bad-path if they result in a 5xx status code.
On average, the LLM generated 6, 11, 7, 8, 10, 10, and 9 scenarios for PetClinic, DayTrader, JPetStore, Feature-service, RestCountries, Genome-Nexus, and App-X, respectively. No scenarios were generated for LanguageTool, which contains only one valid endpoint. Most LLM-generated scenarios span multiple endpoints, with larger models like GPT-o1 generally producing longer sequences—observed in six of the seven applications. Notably, 42.6\% of scenarios span multiple endpoint classes, underscoring the need for more than just isolated endpoint testing.
To illustrate, the following example shows a complex scenario generated by GPT-o1 for Feature-service---the scenario begins by adding a new product (\smalltt{Laptop}), followed by adding features (\smalltt{TouchScreen}, \smalltt{Stylus}), setting constraints, and adding multiple configurations. The complete code for this scenario is available in the supplementary material~\cite{supplementary}.
% \vspace{-15pt}
%<--CRC change
\begin{minted}[frame=lines,framesep=1mm,baselinestretch=1, fontsize=\scriptsize, breaklines, breakanywhere, linenos,numbersep=4pt,]{java}
public class ProductCreationWithFeaturesConstraintsAndConfigurationTest {
    @Test @Order(1)
    void createNewProductLaptop() {...
        .post("/products/Laptop")...}
    @Test @Order(2)
    void addFeatureTouchScreenToLaptop() {...
        .post("/products/Laptop/features/TouchScreen")...}
    @Test @Order(3)
    void addFeatureStylusToLaptop() {...
        .post("/products/Laptop/features/Stylus") ...}
    @Test @Order(4)
    void addFeatureFingerprintScanToLaptop() {...
            .post("/products/Laptop/features/FingerprintScan")...}
    @Test @Order(5)
    void addRequiresConstraintBetweenTouchScreenAndStylus() {...
            .post("/products/Laptop/constraints/requires")...} ...
\end{minted}
% \begin{figure}[t]
%     \centering
%     \includegraphics[width=\linewidth]{figures/testorder.png}
%     \caption*{} % prevents increment
    
% \end{figure}
% \vspace{-5pt}
\begin{findingbox}{3}
{Larger models tend to focus on more complex scenarios in an application, generating scenarios with longer sequences.}
\end{findingbox}

\vskip 2pt
\noindent\textit{Scenario code coverage.} In terms of code coverage, scenario-based test generation does not achieve the same level of performance as individual endpoint testing. While scenario-based test generation is well-suited for validating complex, functional use cases, individual endpoint testing is more effective for maximizing code coverage. 
% We have quantitatively assessed coverage differences, but to understand the representational quality of the test scenarios, we conducted a user study, which will be discussed in the next RQ.
% \begin{findingbox}{3}
% {In terms of code and operation coverages, smaller models perform very similar to their larger counterparts in scenario-based test generation as well.}
% \end{findingbox}
For DayTrader, both code and operation coverage were particularly low during scenario-based test generation. This is largely due to the nature of the application's endpoints—many of which are designed to send pings to various services or perform status checks, making them unsuitable candidates for meaningful scenario construction. 
\begin{findingbox}{4}
{While scenario-based test generation is well-suited for validating complex application use cases, individual endpoint testing is more effective for maximizing code coverage.}
\end{findingbox}
Interestingly, some scenarios extend beyond standard endpoint dependencies. In Genome-Nexus, for example, we observed a cancer hotspot annotation scenario involving three endpoints that are not structurally connected but are semantically related through shared variant and genome location references. This shows that an LLM-based approach can uncover functional links beyond explicit structural dependencies.
\vspace{-8pt}
%<--CRC change
\begin{minted}[frame=lines,framesep=1mm,baselinestretch=1, fontsize=\scriptsize, breaklines, breakanywhere, linenos,numbersep=4pt]{java}

String variant = "7:g.140453136A>T";
Response response = ... .get("/cancer_hotspots/hgvs/{variant}", variant)....
String genomicLocation = "7:140453136";
Response response = ...
        .get("/cancer_hotspots/genomic/{genomicLocation}", genomicLocation)
String[] variants = {"7:g.140453136A>T", "12:g.25398285C>A"};
Response response = ... .post("/cancer_hotspots/hgvs")
\end{minted}
% \begin{figure}[t]
%     \centering
%     \includegraphics[width=\linewidth]{figures/variant.png}
%     \caption*{} % prevents increment
% \end{figure}
\vspace{-25pt}
% Table generated by Excel2LaTeX from sheet 'Sheet8'
\begin{table}[t]
  \centering
  \setlength{\tabcolsep}{3pt}
  \scriptsize
% \vspace{-20pt} 
  \caption{Survey questionnaire.}  
  \resizebox{0.99\linewidth}{!}{
    \begin{tabular}{p{8pt}p{6.5cm}p{0.5cm}}
    \toprule
    \multicolumn{1}{c}{\bf Type} & \multicolumn{1}{c}{\bf Question} & \multicolumn{1}{c}{\bf Format} \\
    \midrule
    \multirow{7}{*}{\begin{sideways}\shortstack{Professional \\Background}\end{sideways}} & {Q1.} Current Role & Open    \\
          & {Q2.} Years of experience in software engineering (incl. education) & MCQ    \\
          & {Q3.} Years of experience in industry & MCQ     \\
          & {Q4.} Level of expertise in Java & MCQ     \\ 
          & {Q5.} Prior experience with automated API-level test generator & MCQ     \\
          & {Q6.} How often do you write API-level tests for your codebase? & MCQ     \\
          & {Q7.} On average, how long do you spend writing API-level tests for an application use case (i.e., tests that exercise a functional scenario)? & MCQ     \\
    \midrule
    \multirow{4}{*}{\begin{sideways}\shortstack{Scenario\\Quality}\end{sideways}} & {Q8.} I understand what this scenario describes & Likert   \\
          & {Q9.} The scenario covers a meaningful functionality of the application & Likert   \\
          & {Q10.} Testing such scenarios is valuable for validating the application & Likert   \\
          & {Q11.} I would test such scenarios if this were my application under test & Likert   \\
    \midrule
    \multirow{8}{*}{\begin{sideways}Generated Test Quality\end{sideways}} & {Q12.} I understand what the test does & Likert   \\
          & {Q13.} The test is well structured & Likert   \\
          & {Q14.} The test feels natural (in terms of variable, method, and class names) & Likert   \\
          & {Q15.} This test correctly implements the test scenario & Likert   \\
          & {Q16.} The input values (e.g., string literals, integer constants), if any, used in the test case are meaningful & Likert   \\
          & {Q17.} The test sequence (i.e., the sequence of API calls) makes sense & Likert   \\
          & {Q18.} The test assertions are meaningful & Likert   \\
          & {Q19.} Would you add the test case to your service-level test suite? & MCQ   \\
    \midrule
     \multirow{1}{*}{\begin{sideways}Comment\end{sideways}} & Q20. What are your thoughts on the strengths and weaknesses of the extracted scenarios and the generated tests? Do you believe they are suitable as outputs from a fully automated process? & Open \\[18pt]
    \bottomrule
    \end{tabular}%
    }
  \label{tb:survey}
  % *5 likert scale has been used
  \vspace{9pt}
\end{table}%

% To understand how often LLMs generate scenarios that go beyond known endpoint dependencies, we manually analyzed each generated scenario and identified those that could not have been constructed solely through producer-consumer or resource-based dependency analysis. We found that ...
% \vspace{-21pt}
\subsubsection{RQ3: Developer perception of scenario-based tests}
We conducted a survey with 41 professional software developers to get qualitative feedback on the test scenarios and corresponding tests generated by \tool.

\vskip 2pt
\noindent\textit{Survey design.} Table~\ref{tb:survey} presents the survey questionnaire, which is broadly divided into four sections.

\vskip 2pt
\noindent\textbf{Professional Background.}
    This section presents seven questions, covering the participants’ professional backgrounds, their roles, experience in software engineering and Java, familiarity with automated API test generators, and time spent in writing API tests.
    % typically spent writing high-quality tests involving multiple API calls.
\vskip 2pt
\noindent\textbf{Test scenario quality.}
    We selected three applications for the user study---PetClinic, Feature-service, and App-X. Participants could select any one of these applications; they were then shown three LLM-generated test scenarios with the corresponding Java test cases. 
    % The scenarios were selected from top-performing model–application pairs (as identified in RQ2): o1 for PetClinic and Feature-service, and DeepSeek-R1-32B for App-X. 
    From multiple \tool runs, three scenarios were randomly sampled per case. Participants were given contextual information about the application and relevant endpoints to aid comprehension. They were then asked to evaluate each scenario for understandability, functional relevance, and inclusion in a test suite.
\vskip 2pt
\noindent\textbf{Generated test quality.} For each test scenario, participants reviewed the corresponding generated code and assessed its clarity and quality. They evaluated whether the test's purpose was clear, the structure logical, and naming conventions (e.g., variables, methods, classes) felt natural. They also judged whether the test accurately implemented the described scenario, whether the API call sequence was meaningful, and whether the assertions were appropriate and relevant.
\vskip 2pt
\noindent \textbf{Overall feedback.} In this section, participants could provide more detailed feedback on \tool, including suggestions for improvement.

\vskip 2pt
\noindent\textit{Participant background.}
The survey participants come from diverse professional roles, including developers, architects, QA engineers, product managers, and researchers. Most had a strong software engineering background—79\% reported over 15 years of experience (including education), and 66\% had more than 10 years in industry. Additionally, 76\% had Java expertise, and 78\% actively performed API testing. Notably, 42\% said writing high-quality API tests takes over 30 minutes. More than half of the participants responded that they have never used any API test-generation tool. These findings highlight both the participants’ qualifications and the potential productivity gains from generating high-quality API tests. More details are available in the supplementary material~\cite{supplementary}.

\begin{figure}[t]
	\centering
		\centering
		\includegraphics[width=\linewidth]{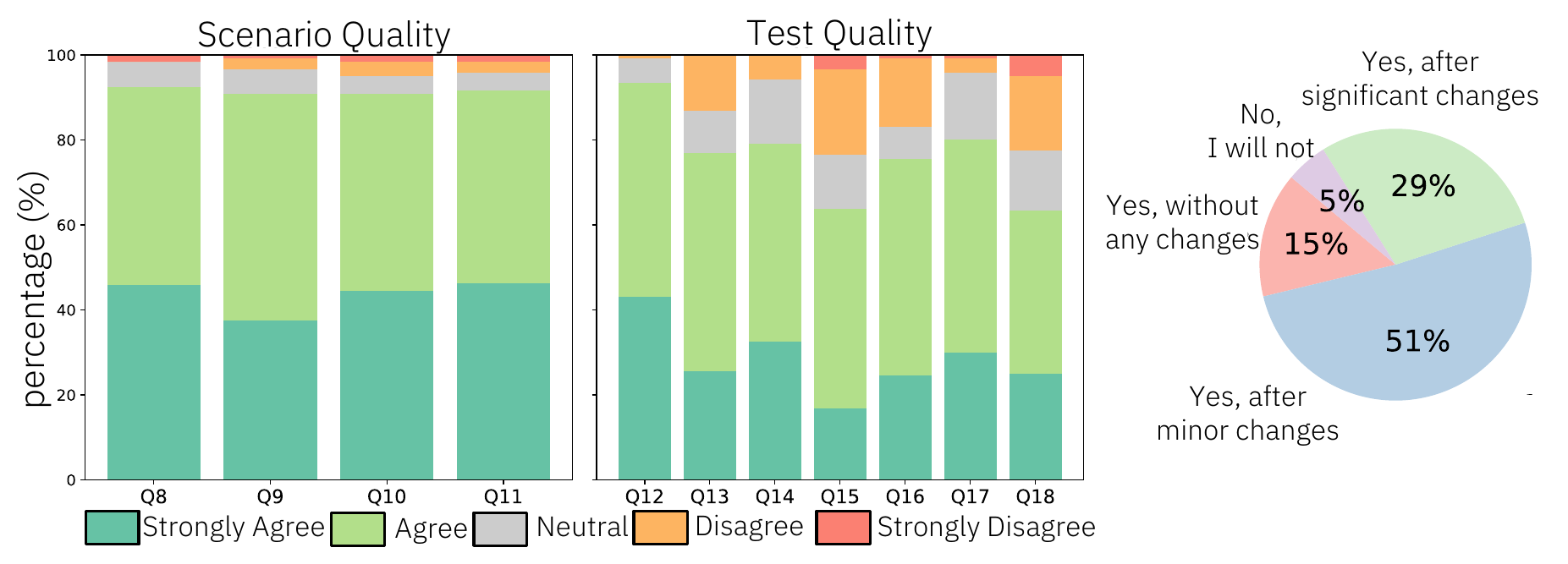}
        %\vspace{-15pt}
\caption{Scenario and test quality assessment (left) and test acceptance (right) by developers.}
\label{fig:test_scenario_quality}
\end{figure}

\begin{findingbox}{5}
{42\% of the participants indicated that writing high-quality API tests is time-consuming, often requiring more than 30 minutes to complete. }
\end{findingbox}
% \begin{figure}
% 	\centering
% 		\centering
% 		\includegraphics[, ,width=0.85\linewidth]{figures/would_you_add.pdf}

% \caption{Q19: Would you add the test case to your service-level test suite?.}
%         \label{fig:q19}
% \end{figure}
% \vskip 2pt
\noindent\textit{Test scenario quality.}
For test scenarios, participants reviewed the natural language descriptions and answered questions about their quality and relevance. The feedback indicated that \tool-generated scenarios were highly rated for understandability, meaningfulness, and alignment with real application use cases. More than 90\% of participants agreed that they would test similar scenarios, while fewer than 5\% expressed a preference against using them.
\begin{findingbox}{6}
{Over 90\% of participants agreed they would test scenarios similar to those automatically extracted by \tool.}
\end{findingbox}
% \vskip 2pt
\noindent\textit{Generated test quality.}
In this phase, developers evaluated the reified test scenarios and the corresponding generated code, focusing on understandability, formatting, and naturalness—particularly in method, class, and variable names, as well as input values. They also assessed the clarity of test sequences, the quality of assertions, and alignment with the described scenarios. More than 70\% of participants found the tests easy to understand and natural. However, over 20\% noted room for improvement in assertion quality and scenario alignment. Still, over 60\% agreed the tests were correctly implemented and included meaningful assertions. Notably, 66\% indicated they would include the generated tests in their suite with minor or no changes, 29\% would do so with significant modifications, and only 5\% would not use them at all (reported in Figure~\ref{fig:test_scenario_quality}).

\begin{findingbox}{7}
{Participants responded positively to various aspects of the generated scenario-based tests, with approximately 66\% indicating that they would add the tests to their regression test suites with little or no modification.}
\end{findingbox}
\vskip 2pt
\noindent\textit{Overall strengths and weaknesses.}
Participant feedback highlighted one of \tool's key strengths: its ability to generate well-structured and readable tests. As one participant noted, ``there are things to make it easier to read … like a central definition of certain variables (\smalltt{BASE\_URI}).'' 
Several participants also suggested improvements to enhance the practicality of \tool. A common request was to make tests more self-contained. Currently, the generated tests rely on hard-coded values and modify the application state without handling setup or cleanup. High-quality tests, however, should create necessary resources at runtime and perform cleanup afterward—an important direction for future development.
Another suggested improvement was enhancing assertion quality. Currently, \tool generates assertions based on response codes and raw server output. Participants noted that validating more success and failure paths—would significantly improve the tests' effectiveness.

\vspace{-5pt}
\subsubsection{RQ4: Effectiveness in triggering failures}
% Table generated by Excel2LaTeX from sheet 'Sheet2'
\begin{table}[!t]
  \centering
  \caption{Server failures triggered.}
  \resizebox{0.96\linewidth}{!}{
    \begin{tabular}{c|r|r|r|r|r|r}
    \toprule
    \multicolumn{1}{c|}{\textbf{Application}} & \multicolumn{1}{c|}{\textbf{Granite-8B}} & \multicolumn{1}{c|}{\textbf{Llama-8B}} & \multicolumn{1}{c|}{\textbf{Devstral}} & \multicolumn{1}{c|}{\textbf{DeepSeek}} & \multicolumn{1}{c|}{\textbf{o1}} & \multicolumn{1}{c}{\textbf{EvoMaster}} \\
    \midrule
    DayTrader & 10    & 2    & 7    & 6   & 2    & - \\
    Spring-PetClinic & 4    & 5   & 5    & 4    & 10    & - \\
    JPetStore & 9    & 7    & 6    & 6    & 6    & - \\
    Feature-service & 76   & 179   & 67    & 65    & 84    & 37 \\
    Restcountries & 1     & 1     & 1     & 1     & 1     &  1\\
    Genome-Nexus & 0     & 0     & 0     & 0     & 0     &  0\\
    Languagetool & 0     & 0     & 0     & 0     & 0     & 6 \\
    App-X  & 0     & 0     & 0     & 0     &     -  & - \\
    \bottomrule
    \end{tabular}%
    }
  \label{tb:faults}%
\vspace{7pt}
\end{table}%

We evaluated \tool{}’s ability to trigger failures using the methodology from prior work~\cite{kim_llamaresttest_2025}. We also reused the regex patterns from a prior work~\cite{martin2022online} to identify unique request–response pairs. To improve generality, we enhanced the implementation so it can parse requests and responses and automatically learn new regex patterns for unseen cases. We found that \tool was able to trigger failures in five of the applications; among the rest, Genome-Nexus and App-X likely did not exhibit failures due to lack of 5xx errors.  \tool triggered more failures compared to EvoMaster on Feature-service but not able to trigger any failure for LanguageTool. On Feature-service, our approach finds many unique failures as the service returns full HTTP pages with stack traces, hindering effective detection via simple string matching.
% To trigger more faults, LLM-based tools like \tool requires adding a feedback loop could improve \tool's fault detection. 

\vspace{-5pt}
\subsubsection{RQ5: Effectiveness of \tool{}'s components}
Our ablation study evaluates the key components of \tool. Due to the high computational demands of testing multiple models, we chose Granite-8B, our smallest effective model and performed the study on four applications with and without OpenAPI specifications: Feature-service, Genome-Nexus, DayTrader, and PetClinic. Table~\ref{tb:ablation} presents the results. Partial orderings from the ODG had the greatest positive effect on test coverage, followed by IPD extraction, value-constraint extraction, the coverage-augmentation agent, and the repair agent.

% Table generated by Excel2LaTeX from sheet 'Sheet6'
\begin{table}[t]
  \centering
  \caption{Results of the ablation study.}
  \resizebox{.9\linewidth}{!}{
    \begin{tabular}{c|l|rrrrr}
    \toprule
          \multicolumn{2}{c|}{\shortstack{Coverage Metric}}    & \multicolumn{1}{l}{\shortstack{Value\\Constraints}} & \multicolumn{1}{l}{IPD} & \multicolumn{1}{l}{\shortstack{Partial\\Order}} & \multicolumn{1}{l}{\shortstack{Request\\Fixing}} & \multicolumn{1}{l}{\shortstack{Coverage\\Augmentation}} \\
   \midrule
    \multirow{3}[0]{*}{Application} & Line  & +9.0     & +12.0    & +17.1  & +2.2   & +0.4 \\
    
          & Branch & +33.2  & +26.0    & +34.2  & +8.4   & +13.6 \\
          & Database & +7.2   & +35.0    & +14.0    & -1.3  & +5.2 \\
    \midrule
    \multirow{3}[0]{*}{Reachability} & Line  & +15.4  & +17.0    & +27.0    & +5.5   & +1.9 \\
          & Branch & +31.7  & +31.0    & +36.6  & +12.8  & +14.9 \\
          & Database & +5.7   & +38.0    & +16.3  & -1.2  & +6.3 \\
    \bottomrule
    \end{tabular}%
}
  \label{tb:ablation}%
  \vspace{5pt}
\end{table}%

\vspace{-5pt}
\paragraph{Cost of using \tool} A major concern with LLM-based approaches is their operational cost, as excessive token consumption can substantially increase overall expense and limit practical usage. To evaluate this aspect, we measured the token usage for both endpoint-based and scenario-based test generation, and translated these values into monetary cost using the pricing model of different LLMs. Our analysis demonstrates that \tool is significantly cost-efficient, both in terms of total cost and the number of LLM invocations. Specifically, for Devstral (via OpenRouter~\cite{openrouter}), the average cost is \$0.24 for endpoint-focused and \$0.22 for scenario-based generation, whereas for GPT-o1 (on the OpenAI platform), the cost is \$6.17 and \$4.42, respectively. The detailed results are included in our supplementary material~\cite{supplementary}. 
% We believe this efficiency primarily because of the integration of static analysis, which enables \tool to perform targeted pre-processing and invoke the LLM only when necessary, rather than feeding the entire application context to an off-the-shelf model.

% \input{discussion}
\section{Related Work}
\label{sec:related}

% Automated REST API testing techniques are categorized into black-box and white-box approaches~\cite{martin2021:blackandwhite, nidhra2012black:blackandwhite}; former relies on API specs and the latter on source code inspection and runtime monitoring.

\vskip 2pt
\noindent\textit{Black-box API testing.}
Early black-box techniques use fuzzing and model-based strategies to generate request sequences from the OpenAPI Specification. Tools such as RESTler~\cite{atlidakis_restler_2019}, RestTestGen~\cite{viglianisi_resttestgen_2020}, MoREST~\cite{liu_morest_2022}, and RAFT~\cite{saha:2025:raft} conduct stateful fuzzing via graph traversals and HTTP method differentiation. Recent LLM-enhanced tools, like KAT~\cite{le_kat_2024} and LogiAgent~\cite{zhang_logiagent_2025}, infer semantic relationships, outperforming heuristic methods. Other LLM-augmented API testing systems, including RESTGPT~\cite{kim_leveraging_2024}, AutoRestTest~\cite{kim_multi-agent_2025, stennett_autoresttest_2025}, and LlamaRestTest~\cite{kim_llamaresttest_2025}, improve parameter generation and IPD extraction. AutoRestTest and LogiAgent also use agents for iterative test refinement. However, these methods are limited by inaccuracies in OpenAPI specifications~\cite{deng2025lrasgen, martin2021specification}.

\vskip 2pt
\noindent\textit{White-box API testing.}
White-box techniques use internal system knowledge to enhance coverage and fault detection. EvoMaster~\cite{arcuri_evomaster_2018} applies evolutionary algorithms based on code coverage and mutations, while MioHint~\cite{li_llm-assisted_2025} uses LLMs to improve inputs for challenging branches through static analysis. 
% However, both face difficulties in creating semantically meaningful request sequences.

\vskip 2pt
\noindent\textit{Scenario-based testing.}
Scenario-based testing simulates real-world workflows through dependent API sequences. Traditional black-box tools, such as RESTler~\cite{atlidakis_restler_2019}, MoREST~\cite{liu_morest_2022}, and RestTestGen~\cite{viglianisi_resttestgen_2020}, infer dependencies heuristically. Recent tools like RAFT~\cite{saha:2025:raft}, KAT~\cite{le_kat_2024}, and LogiAgent~\cite{zhang_logiagent_2025} integrate LLMs, with LogiAgent generating scenarios from endpoint descriptions. However, these approaches rely solely on OpenAPI specs and overlook code and database logic.

\noindent
\textit{REST API specification generation.} Several techniques have been proposed to automatically generate REST API specifications for test generation. Commercial tools like SpringFox~\cite{springfox}, springdoc-openapi~\cite{springdoc}, and Swagger Core~\cite{swaggerCore} generate OpenAPI specs via runtime inspection but are framework-specific and often incomplete, requiring manual effort for integration. Research tools such as Respector~\cite{huang:2024:respector} produce higher-quality specs but supports only Spring and JAX-RS applications on specific Java versions.
% In contrast, \tool requires no OpenAPI specification and goes beyond testing individual endpoints.

\vskip 2pt
\noindent\textit{Positioning \tool.}
\tool is the first white-box, LLM-based agentic framework for REST API testing. It extends model-based techniques (e.g., RESTler, RestTestGen, KAT) by leveraging static analysis to identify dependencies and sequence operations beyond the OpenAPI specification, enabling accurate, meaningful scenarios. 
% End-to-end LLM integration generates semantic tests that surpass fuzzing-based white-box methods like EvoMaster. 
Compared to LogiAgent---the closest counterpart---\tool incorporates code understanding for request sequencing and generation, overcoming reliance on incomplete external resources. Moreover, \tool's comprehensive tool-chain for static analysis, IPD extraction, and generation enhances autonomy and adaptive reasoning. 
\section{Threats to Validity}
\label{sec:threats}
% This section reports the potential threats to validity of the key findings of this paper:
% \be
% \item \textit{Construct Validity:~} 
% To evaluate \tool, we used line and branch coverage with server-side error detection. However, these metrics may not fully capture the application's behavioral and business validity. Therefore, our test generation includes scenario-based tests that replicate functional workflows. We used both quantitative and qualitative measures to assess the quality of these scenarios, including developer feedback via a survey.

The use of LLMs and agents derived from LLMs exposes \tool to an inherent stochasticity and sensitivity to prompt structure that may affect repeatability of our results. To limit this, we used temperature as low as 0.2 and performed two runs with each LLM. However, to mitigate the impact of running fewer trials on the reported coverage, we repeated the experiment with Devstral (due to its lower inference cost) ten times. We found that the standard deviation of both line and branch coverage was very low (for branch coverage, 0.0–5.2\% for individual endpoints and 0.0–7.6\% for scenario-based tests; for line coverage, 0.1–1.3\% for individual endpoints and 1.4–5.1\% for scenario-based tests). These results are available in our artifact \cite{supplementary}.
% Further, agents for coverage augmentation and repair may subtly alter the test structure. We therefore evaluate our tests with developers (via our survey) to ensure overall correctness and usefulness of the generated tests.

Our evaluation covered eight applications built on various frameworks, both with and without OpenAPI specifications. We focused on synchronous REST APIs in Java, so the results may not generalize to applications using other communication models (e.g., message queues or event-driven architectures) or implemented in other languages. Like EvoMaster, \tool currently cannot mock external services—a limitation we plan to address in future work.
% \ee
\section{Conclusion}
\label{sec:conclusion}
We presented \tool, a white-box approach for service-level testing of enterprise Java applications combining static analysis with LLM-driven agentic workflows. \tool generates both endpoint-focused and scenario-based test cases, targeting high code coverage and realistic use-case execution. Our technique integrates symbolic information extracted through static analysis with semantic reasoning capabilities of LLMs, supported by agents that repair, augment, and compose tests. Evaluation across eight applications demonstrates that \tool outperforms prior approaches in coverage and scenario realism, with favorable feedback from developers and supporting ablation analyses.
We identify several future research directions: extending \tool to support more backends (e.g., Python Flask, Node.js Express) and interfaces (e.g., GraphQL, gRPC), improving the generated scenario-based tests to be self-contained, and developing human-in-the-loop variants of \tool for interactive scenario generation and domain-specific test refinement.
% enhancing CI/CD workflows, prioritizing tests through change-based analysis,

% \balance
\bibliographystyle{ACM-Reference-Format}
\bibliography{references}

\end{document}